\documentclass[preprint,onecolumn,nofootinbib]{revtex4}
\usepackage[colorlinks=true,linkcolor=blue,urlcolor=blue,filecolor=black,citecolor=red,pdfstartview=FitV,pdftitle={},pdfsubject={},pdfkeywords={},pdfpagemode=None,bookmarksopen=true]{hyperref}
\usepackage{graphicx}
\usepackage{amsmath}
\usepackage{amsfonts}
\usepackage{amssymb,ulem}
\usepackage{color,xcolor}%
\usepackage{CJK}
\usepackage{subfigure}
\usepackage{amsthm,amsmath,amssymb}
\usepackage{mathrsfs}
\usepackage{multirow}

\usepackage{dcolumn}
\usepackage{float}
\begin{document}
\title{Observational appearances of an inner extremal regular black hole illuminated by various accretion flows}
\author{Dan Zhang $^{1}$}
\thanks{danzhanglnk@163.com}
\author{Guoyang Fu$^{2}$}
\thanks{FuguoyangEDU@163.com}
\author{Xi-Jing Wang$^{3}$}
\thanks{xijingwang01@163.com}
\author{Qiyuan Pan$^{4}$}
\thanks{panqiyuan@hunnu.edu.cn}
\author{Xiao-Mei Kuang$^{1}$}
\thanks{xmeikuang@yzu.edu.cn}
\author{Jian-Pin Wu$^{1}$}
\thanks{jianpinwu@yzu.edu.cn}
\affiliation{$^1$\mbox{Center for Gravitation and Cosmology, College of Physical Science and Technology,} \mbox{Yangzhou University, Yangzhou 225009, China}\\
$^2$\mbox{Department of Physics and Astronomy,
Shanghai Jiao Tong University,}\mbox{Shanghai 200240, China}\\
$^3$~ \mbox{Department of Astronomy, School of Physics and Technology, Wuhan University,} \mbox{Wuhan 430072, China}\\$^4$\mbox{Department of Physics, Key Laboratory of Low Dimensional Quantum }  \mbox{ Structures and Quantum Control of Ministry of Education,Institute of Interdisciplinary } \mbox{Studies, and Synergetic Innovation Center for Quantum Effects and Applications,} \mbox{Hunan Normal University,  Changsha, Hunan 410081, China}
}

\begin{abstract}

This paper investigates the observational appearances of an inner extremal regular black hole (IERBH) illuminated by various types of accretion models. The study reveals that when the BH is illuminated by specific accretion flows, the effects of quantum gravity become more pronounced, significantly impacting key observational features such as the shadow radius, photon ring, and total observed intensity. Specifically, the introduction of a more realistic radially infalling spherical accretion flow further accentuates these differences. This dynamic flow results in a darker central region in the BH image due to the Doppler effect, which modulates the observed intensity based on the relative motion of the infalling matter. The shadow radius and total observed intensity are notably affected by the quantum correction parameters, providing additional signatures that distinguish regular BHs from their classical counterparts.

\end{abstract}

\maketitle
\tableofcontents

\section{Introduction}\label{sec-intro}

Observing black holes (BHs) is a challenging task because of their strong gravitational fields, which hinder light from escaping. Notwithstanding these challenges, the Event Horizon Telescope (EHT) collaboration has successfully captured ultra-high angular resolution pictures of the supermassive BHs in M87* \cite{EventHorizonTelescope:2019dse,EventHorizonTelescope:2019uob,EventHorizonTelescope:2019jan,EventHorizonTelescope:2019ths,EventHorizonTelescope:2019pgp,EventHorizonTelescope:2019ggy}, and in Sgr A*, located at the center of the Milky Way galaxy \cite{EventHorizonTelescope:2022wkp,EventHorizonTelescope:2022apq,EventHorizonTelescope:2022wok,EventHorizonTelescope:2022exc,EventHorizonTelescope:2022urf,EventHorizonTelescope:2022xqj}, achieving a groundbreaking milestone. 

BH images provide valuable information about BH shadows and photon rings \cite{Synge:1966okc,Falcke:1999pj,Mishra:2019trb,Amir:2016cen,Huang:2016qnl,Grenzebach:2014fha,Zhang:2022klr,Chen:2023wzv,Zhang:2024jrw,Liu:2024soc,Kuang:2024ugn}. Additionally, the lights emitted by nearby sources significantly influence optical features, particularly the BH's shadow. A given accretion disk surrounding a BH can serve as one of the light sources, which can be used to extract more properties of the central BH.
In 1979, Luminet pioneered the study of images portraying geometrically thin accretion disks around Schwarzschild BH \cite{Luminet:1979nyg}. It is hypothesized that thin disks mitigate the influence of the accretion disk's mass on the spacetime background, enabling close examination of the lensing ring or photon ring effect. More recently, Samuel E. Gralla et al. studied the images of Schwarzschild BH with thin and thick accretion disks, finding that the photon ring and lensing ring significantly contribute to the BH shadows and the corresponding observed flux \cite{Gralla:2019xty}. They also proposed categorizing light trajectories in the vicinity of a BH into direct, lensed ring and photon ring emissions, to better investigate how these different emissions affect the optical appearance of the BH. Furthermore, researchers have found that the size of the BH shadow depends on the geometry of spacetime in the neighborhood of the event horizon, facilitating the detection of quantum gravity effects and providing insights into how quantum BHs deviate from classical BHs \cite{Peng:2020wun,He:2021htq,Zeng:2023fqy}. Building on these advancements, numerous studies have been carried out to investigate the effects of modified gravity on the optical appearances of BHs \cite{Bambi:2013nla,Zeng:2020dco,Guo:2021bwr,Zeng:2021mok,Wang:2023vcv,Saurabh:2020zqg,Qin:2020xzu,Gan:2021pwu,Meng:2023htc,Okyay:2021nnh,Li:2021ypw,Li:2021riw,Guo:2021bhr,Wen:2022hkv,Chakhchi:2022fls,Hou:2022eev,Kuang:2022xjp,Uniyal:2022vdu,Meng:2024puu,Uniyal:2023inx,Yang:2024utv,Wang:2024lte,Li:2024kyv}.

Recently, regular BHs have been widely studied. These models resolve the singularity problem of spacetime by replacing the central singularity with a non-singular core. Generic models of regular BHs typically feature two horizons: the inner horizon and the outer horizon \cite{Carballo-Rubio:2019fnb,Bonanno:2020fgp}. In 1968, Penrose observed that the radiative tail of a collapsing BH undergoes an exponential blueshift near the inner horizon, indicating its high instability against time-dependent external perturbations \cite{Penrose:1968r,Simpson:1973ua}. Subsequent investigations into the dynamics near the inner horizon revealed an exponential instability, a phenomenon known as mass inflation instability \cite{Poisson:1989zz,Ori:1991zz,Hamilton:2008zz}. This mass inflation instability raises significant questions about the physical viability of regular BHs as alternatives to their singular counterparts.
To circumvent the classical mass inflation instability, a novel regular BH with vanishing surface gravity at the inner horizon was proposed in \cite{Carballo-Rubio:2022kad}.
We refer to this model as an inner extremal regular BH (IERBH). This type of regular BH features a stable core and can effectively replace conventional regular BHs, thereby enhancing the study of phenomenology. Related studies can be found in \cite{Carballo-Rubio:2022twq,Franzin:2022wai,PhysRevD.108.128501,PhysRevD.107.024005,McMaken:2023uue,Ghosh:2022gka}.

In \cite{Cao:2023par}, the authors investigated the shadow and optical appearance of the IERBH proposed in \cite{Carballo-Rubio:2022kad} and discovered that when a photon is emitted and received within the same universe, the observed optical image shows only slight differences from that of a Schwarzschild BH. This finding indicates that detecting the effects of quantum gravity is highly challenging. However, the aforementioned work primarily focuses on the optical images of BHs illuminated by the thin accretion disk, where the radiation originates from the photon sphere. As is widely understood, the BH image depends not only on the BH geometry but also on the characteristics of the accretion models. Whether other accretion models exhibit significant differences in their optical images and how these differences may impact the feasibility of probing quantum gravity effects remain subjects for further investigation and discussion. In this paper, we will further investigate various types of accretion models illuminating this IERBH to examine how different accretion models affect the BH images and to explore the detectability of quantum gravity effects in the optical images of BHs.

The organization of this work is as follows. In Section \ref{sec-wmbh}, we provide a concise overview of the geometry of IERBH. In Section \ref{tra-pho}, the photon trajectories around the IERBH were then calculated with the help of the Lagrangian formalism and ray-tracing method. Additionally, we analyze the impact of the quantum correction parameters $a_2$ and $b_2$ on the photon trajectory and the critical impact parameter. Next, we investigate the shadows, photon rings, and corresponding observed intensities of IERBHs illuminated by thin disk accretion in Section \ref{rin-img} and thin spherical accretions flow in Section \ref{spherical-img}, respectively. We summarize our findings in Section \ref{conslusion}. In Appendix \ref{method}, we provide a brief overview of the observational characteristics of thin accretion disks.

\section{An inner extremal regular black hole}\label{sec-wmbh}

We start from an IERBH characterized by the following metric \cite{Carballo-Rubio:2022kad}
\begin{eqnarray} \label{metric}
	&&
	ds^2=-F(r)dt^2+\frac{1}{F(r)}dr^2+r^2(d\theta^2+\sin^2\theta d\phi^2)\,,
\end{eqnarray}
with
\begin{eqnarray} \label{fr}
	F(r)=\frac{(r-r_-)^3(r-r_+)}{(r-r_-)^3(r-r_+)+2Mr^3+(a_2-3r_-(r_++r_-))r^2}\,.
\end{eqnarray}
Here, $r_-$ and $r_+$ denote the inner and outer horizons, respectively, which are given by:
\begin{eqnarray} \label{rfrz1}
	&&
	r_{-}=\frac{1}{12}\left(-6M+3\delta+\sqrt{3}\sqrt{8a_2-8b_2+3(-2M+\delta)^2}\right)\,,\nonumber
	\\
	&&
	r_{+}=\frac{1}{12}\left(18M-9\delta+\sqrt{3}\sqrt{8a_2-8b_2+3(-2M+\delta)^2}\right)\,,
\end{eqnarray}
where $a_2$, $b_2$ and $\delta$ are the quantum parameters that signify deviations from the Schwarzschild BH. The introduction of the quantum corrections in this novel BH \eqref{metric} with Eq.\eqref{fr} and Eq.\eqref{rfrz1} successfully avoids the singularity at the BH center \cite{Carballo-Rubio:2022kad}. Notably, the surface gravity of the inner horizon of this model vanishes, effectively preventing the mass inflation instability \cite{Carballo-Rubio:2022kad}. For simplicity, throughout this paper, we only focus on the case where $\delta=0$, then Eq.\eqref{rfrz1} is simplified as
\begin{eqnarray} \label{rfrz}
	&&
	r_{-}=\frac{1}{6}(-3M+\sqrt{6a_2-6b_2+9M^2})\,,\nonumber
	\\
	&&
	r_{+}=\frac{1}{6}(9M+\sqrt{6a_2-6b_2+9M^2})\,.
\end{eqnarray}
Throughout this paper, we will work with dimensionless quantities. To achieve this, we rescale all relevant physical quantities, including the parameters $a_2$, $b_2$, and the impact parameter $b$ discussed below, by the BH mass $M$. This is equivalent to set $M=1$.

\begin{figure}[ht]
	\centering
	\subfigure{
		\includegraphics[width=8cm]{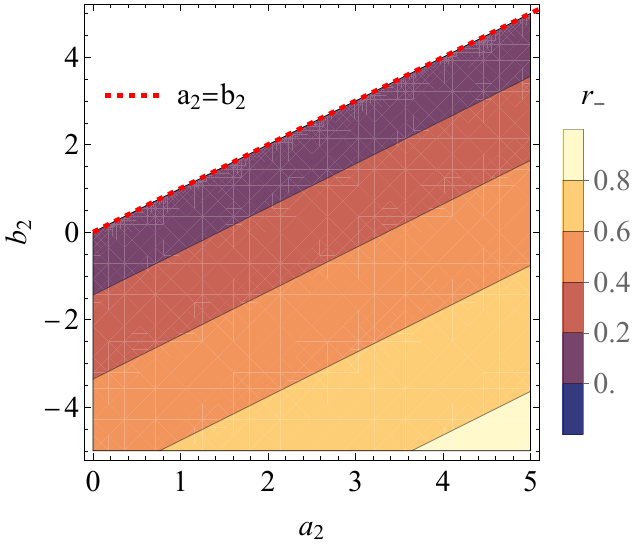}}
	\subfigure{
		\includegraphics[width=8cm]{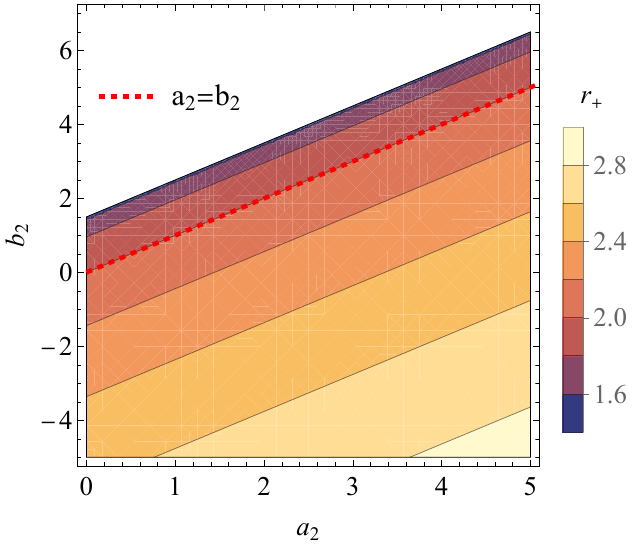}}
	\caption{The inner horizon $r_-$ (left) and outer horizon $r_+$ (right) as the functions of the $a_2$ and $b_2$. The red dashed line represents $a_2=b_2$.}
	\label{horizon_stru}
\end{figure}

\begin{figure}[ht]
	\centering
	\subfigure{
		\includegraphics[width=8.5cm]{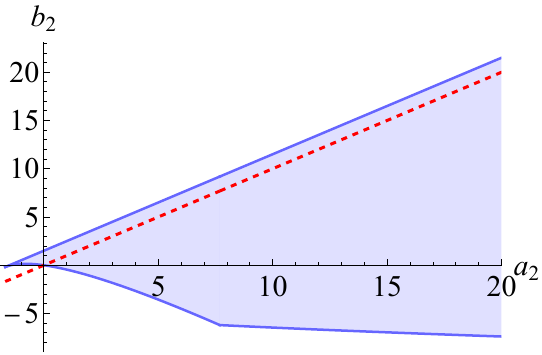}}
	\caption{The allowed range of the parameters $a_2$ and $b_2$. The red dashed line represents $a_2=b_2$.}
	\label{a2b2range}
\end{figure}

Fig.\ref{horizon_stru} shows the inner horizon $r_-$ and the outer horizon $r_+$ as the functions of the parameters $a_2$ and $b_2$. When $a_2=b_2=0$, the IERBH described by Eq.\eqref{metric} reduces to the Schwarzschild BH. For $a_2=b_2\neq 0$, this BH retains the same horizon structure as the Schwarzschild BH. From Fig.\ref{horizon_stru}, it is evident that the inner horizon vanishes when $a_2>b_2$, suggesting that this IERBH possesses only a single horizon $r_+$. Conversely, when $a_2<b_2$, both the inner horizon $r_-$ and the outer horizon $r_+$ are present, indicating that this IERBH possesses double horizons. Finally, we present the allowed range of the parameters $a_2$ and $b_2$, as illustrated in Fig.\ref{a2b2range}. 

\section{Trajectories of photons around IERBH}\label{tra-pho}

In this section, we analyze the null geodesics around this IERBH. The corresponding Lagrangian of photon is given by
\begin{eqnarray}\label{math-L0}
	\mathcal{L}=\frac{1}{2}g_{\mu \nu}(X)\dot{X}^{\mu}\dot{X}^{\nu}\,,
\end{eqnarray}
where the dot represents the derivative with respect to the affine parameter $\lambda$, and the four velocity of the photon is defined as $\dot{X}^\mu=d X^{\mu}/d\lambda$. By performing a variation, we can derive two conserved quantities associated with the motion of the photon: the energy $E$ and the angular momentum $L_z$. These quantities are worked out as follows:
\begin{eqnarray}\label{ELz}
	E\equiv -\frac{\partial \mathcal{L}}{\partial \dot{t}}=F(r)\dot{t}\,,~~~ L_z\equiv \frac{\partial \mathcal{L}}{\partial \dot{\phi}}=r^2\dot{\phi}\,.
\end{eqnarray}
In \cite{Luminet:1979nyg}, Luminet states that the motion of a photon is not determined by its energy $E$ and angular momentum $L_z$ individually, but rather by their ratio. In the subsequent discussion, we define the ratio $b\equiv L_z/E$ as the impact parameter.

By redefining the affine parameter $\lambda$ as $\lambda/L_z$, we obtain the following formula for the four velocities in this spherically symmetric spacetime:
\begin{eqnarray}\label{four-vel}
&&
\dot{t}=\frac{1}{bF(r)}\,,\label{td}\\
&&
\dot{\phi}=\pm\frac{1}{r^2}\,,\label{phid}\\
&&
\dot{r}^2=\frac{1}{b^2}-V_{\text{eff}}(r)\,,
\label{rdot}
\end{eqnarray}
where the symbol ``$\pm$" denotes the direction of the photon's travel in the equatorial plane, with ``$+$" for counterclockwise and ``$-$" for clockwise. The effective potential $V_{\text{eff}}$ for the photon is read as 
\begin{eqnarray}\label{Veffd}
	V_{\text{eff}}(r)=\frac{F(r)}{r^2}\,.
	\label{Veffd}
\end{eqnarray}

Before investigating the trajectory of a photon, we require that the photon sphere satisfies circular orbit conditions, characterized by $\dot{r}=0$ and $\ddot{r}=0$. Recalling Eq.\eqref{rdot}, we can concretely derive the conditions for the photon sphere as follows:
\begin{eqnarray}
	V_{\text{eff}}(r)|_{r=r_{ph}}=\frac{1}{b_{ph}^2}\,,~~~ V'_{\text{eff}}(r)|_{r=r_{ph}}=0\,,
	\label{Veffdd}
\end{eqnarray}
where $r_{ph}$ and $b_{ph}$ are the radius and critical impact parameter of the photon sphere, respectively. 

\begin{figure}[ht]
	\centering
	\includegraphics[width=5.2cm]{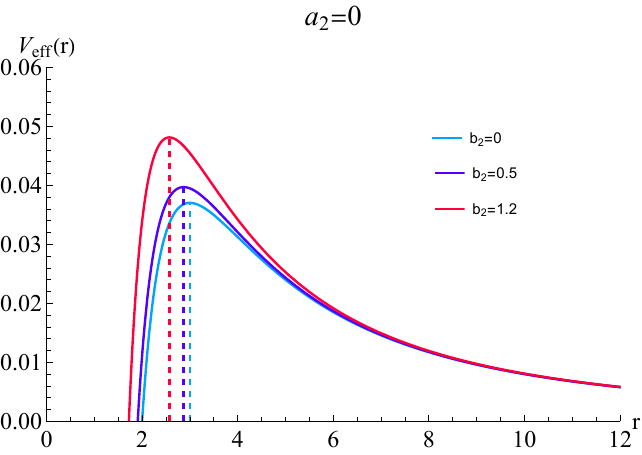}
	\includegraphics[width=5.2cm]{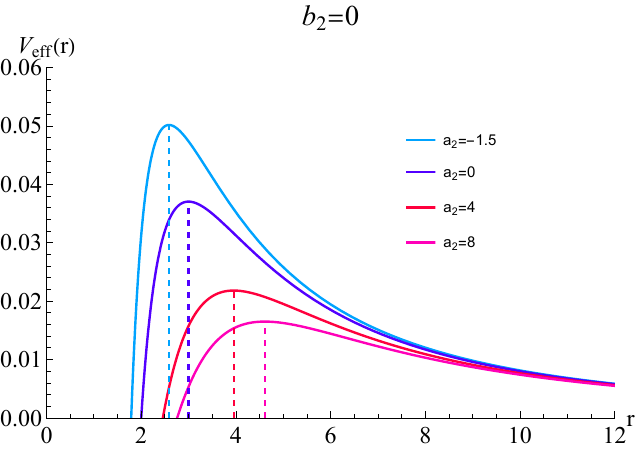}
	\includegraphics[width=5.2cm]{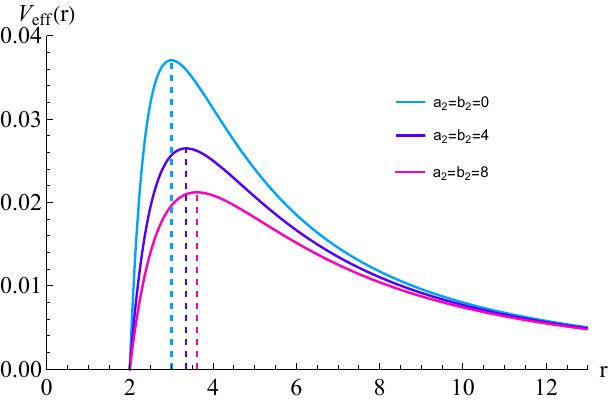}
	\caption{The effective potential $V_{\text{eff}}$ as a function of the radial coordinate $r$ for various values of the parameters $a_2$ and $b_2$.}
	\label{Veff}
\end{figure}

Fig.\ref{Veff} shows how the effective potential $V_{\text{eff}}$ changes with different parameters $a_2$ and $b_2$. We first turn off the parameter $a_2$ and release $b_2$. As $b_2$ increases from zero, the peak of the effective potential $V_{\text{eff}}$ rises, while the position of the peak moves to the left, as shown in the left plot of Fig.\ref{Veff}. It means that both the radius of the photon sphere $r_{ph}$ and the critical impact parameter $b_{ph}$ decrease as $b_2$ increases, as also presented in Table \ref{tab_regular_v1}. In contrast, the parameter $a_2$ exhibits a fully opposite effect on the behaviors of the effective potential $V_{\text{eff}}$, as depicted in the middle plot of Fig.\ref{Veff}. Specifically, an increase in $a_2$ leads to larger $r_{ph}$ and $b_{ph}$ (see the middle plot of Fig.\ref{Veff} and also see the Table \ref{tab_regular_v3}).

\begin{table}[H]
	\centering
	\begin{tabular}{|c|cc|cc|cc|cc|c| }
		\hline 
		& $b_2=0$   &~& $b_2=0.5$  &~& $b_2=1$  &~& $b_2=1.2$ &~& $b_2=1.5$ \\ 
		\hline
		$r_{ph}$ & 3.00000  &~&  2.86969 &~& 2.68045  &~& 2.57062 &~& 2.17143\\
		\hline
		$b_{ph}$ & 5.19615  &~&  5.01801 &~& 4.73406  &~& 4.55793 &~& 3.85526\\
		\hline
	\end{tabular}\\
	\caption{The radius of photon sphere $r_{ph}$, and the impact parameter $b_{ph}$ for various values of $b_2$. Here, we have fixed $a_2=0$.
	}  \label{tab_regular_v1}
\end{table}

Next, we consider the case where $a_2=b_2$, resulting in the horizon of this IERBH aligning with that of the Schwarzschild BH. We show the results of $V_{\text{eff}}$ in the right plot of Fig.\ref{Veff}. We can observe that, as the parameters $a_2$ and $b_2$ increase, the peak of the effective potential is suppressed, indicating that the influence of $a_2$ is dominant over that of $b_2$. Moreover, as the parameters $a_2$ and $b_2$ increase, both $r_{ph}$ and $b_{ph}$ also increase (see Table \ref{tab_regular_v5}), further demonstrating that the influence of $a_2$ is more significant. 

\begin{table}[H]
	\centering
	\begin{tabular}{|c|cc|cc|cc|cc|c| }
		\hline 
		& $a_2=-1.5$  &~&   $a_2=2$   &~& $a_2=4$  &~& $a_2=6$ &~& $a_2=8$ \\ 
		\hline
		$r_{ph}$ & 2.06291   &~&  3.54155  &~&  3.95378 &~& 4.30228 &~& 4.61015\\
		\hline
		$b_{ph}$ & 3.48431  &~&  6.10779  &~&  6.77102 &~& 7.31505 &~& 7.78511 \\
		\hline
	\end{tabular}\\
	\caption{The radius of photon sphere $r_{ph}$, and the impact parameter $b_{ph}$ for various values of $a_2$. Here, we have fixed $b_2=0$.
	}  \label{tab_regular_v3}
\end{table}
\begin{table}[H]
	\centering
	\begin{tabular}{|c|cc|cc|cc|cc|c| }
		\hline 
		& $a_2=b_2=0$ &~& $a_2=b_2=2$ &~& $a_2=b_2=4$  &~& $a_2=b_2=6$  &~& $a_2=b_2=8$\\ 
		\hline
		$r_{ph}$ & $3.00000 $ &~& 3.19582 &~& 3.35530 &~& 3.49203 &~& 3.61289\\
		\hline
		$b_{ph}$ & $5.19615$ &~& 5.71313 &~& 6.14607 &~& 6.52556 &~& 6.86723\\
		\hline
	\end{tabular}\\
	\caption{The radius of photon sphere $r_{ph}$, and the impact parameter $b_{ph}$ for various values of $a_2=b_2$. 
	}  \label{tab_regular_v5}
\end{table}

We now turn our attention to the trajectories of photons around the BH, which is governed by the following equation:
\begin{eqnarray}
	\frac{dr}{d\phi}=\pm r^2\sqrt{\frac{1}{b^2}-V_{\text{eff}}(r)}\,.
	\label{rphi}
\end{eqnarray}
This compact equation is derived by combining Eqs.~\eqref{phid} and \eqref{rdot}. It is evident that given the BH metric, the impact parameter $b$ plays a decisive role in determining the trajectory of the photon. Specifically, in the region where $b>b_{ph}$, photons coming from infinity encounter a potential barrier, pass through the turning point and then are scattered into infinity. At $b=b_{ph}$, light travels in a circular motion with a non-zero angular velocity close to the photon sphere, allowing it to circle the BH infinitely. Finally, in the region where $b<b_{ph}$, with no potential barriers to block it, the photon will be captured by the BH and eventually fall into it.

\section{Images of IERBH illuminated by disk accretions}\label{rin-img}

In this section, we will investigate the observed shadow images, rings, and optical appearance of this IERBH surrounded by various accretion disk scenarios. We follow the setup described in \cite{Gralla:2019xty}, where the accretion disk is located on the equatorial plane of the BH, and the observer is positioned at the North Pole.

\subsection{Classification of rays: direct, lensed ring and photon ring emissions}

Considering an accretion disk situated on the equatorial plane of the BH, we can classify the light rays based on the orbit number $n$ of the photon defined by $n=\phi/(2\pi)$, where $\phi$ represents the total change in the azimuthal angle of the photon as it orbits the black hole. This classification, as outlined in \cite{Gralla:2019xty}, results in three distinct categories:
\begin{enumerate} 
	\item Direct emission ($n < 3/4$): Light rays that intersect the accretion disk only once.
	\item Lensed ring emission ($3/4 < n < 5/4$): Light rays that intersect the accretion disk twice.
	\item Photon ring emission ($n > 5/4)$: Light rays that intersect the accretion disk three or more times.
\end{enumerate} 
For a more comprehensive understanding of the three classes of trajectories described above, readers are referred to \cite{Wielgus:2021peu,Bisnovatyi-Kogan:2022ujt} for schematic diagrams that illustrate these emissions in detail.

\begin{figure}[ht]
	\centering
	\includegraphics[width=5.2cm]{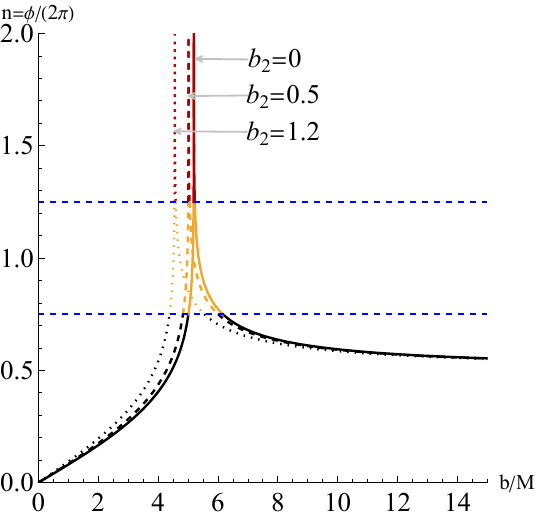}
	\includegraphics[width=5.2cm]{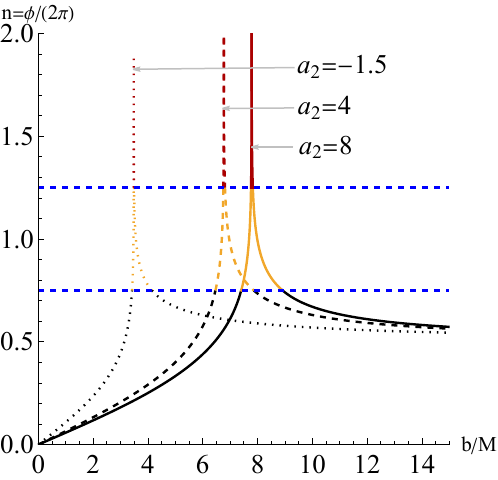}
	\includegraphics[width=5.2cm]{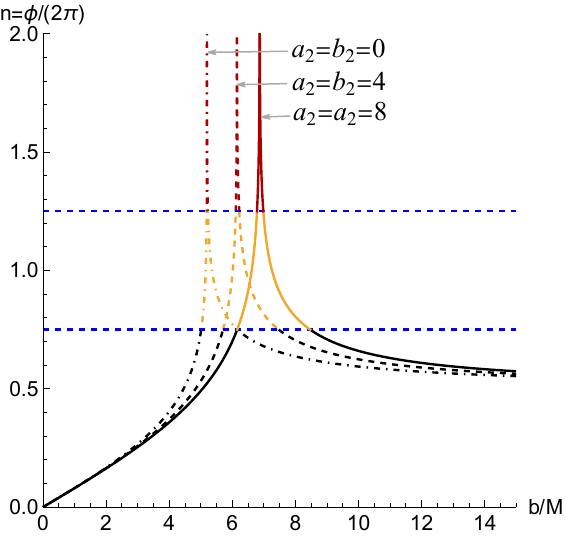}
	\caption{The orbit number $n$ as a functions of the impact parameter $b$ for different values of $a_2$ and $b_2$. The blue dashed line corresponds to $n=3/4$ and $n=5/4$. The direct $(n<3/4)$, lensed ring $(3/4 <n<5/4)$, and photon ring $(n>5/4)$ emissions correspond to the black, yellow, and red curves, respectively.}
	\label{totaln}
\end{figure}
\begin{table}[H]
	\centering
	\begin{tabular}{|c|cc|cc|c| }
		\hline 
		$b_2$ & $Direct(n<3/4)$   &~& $Lensed$ $Ring(3/4<n<5/4)$  &~& $Photon$ $Ring(n>5/4)$ \\ 
		\hline 
		0 & $b<5.01514$ $\|$ $b>6.16757$ &~&  $5.01514<b<5.18781$ $\|$ $5.22794<b<6.16757$ &~&$5.18781<b<5.22794$\\
		\hline
		0.5 &  $b<4.83241$ $\|$ $b>6.00643$ &~&  $4.83241<b<5.00899$ $\|$ $5.05182<b<6.00643$ &~& $5.00899<b<5.05182$\\
		\hline
		1.2 & $b<4.38480$ $\|$ $b> 5.54596$ &~&  $4.38480<b<4.54932$ $\|$ $4.59216<b<5.54596$ &~& $4.54932<b<4.59216$\\
		\hline
		1.5 & $b<3.72967$ $\|$ $b> 4.75598$ &~&  $3.72967<b<3.84989$ $\|$ $3.88191<b<4.75598$ &~& $3.84989<b<3.88191$\\
		\hline
	\end{tabular}\\
	\caption{The ranges of the impact parameter $b$ corresponding to the direct emission, lensed ring emission, and photon ring emission of the regular BH, given $a_2=0$, for various values of $b_2$.
	}
	\label{tab_regular_v2}
\end{table} 
\begin{table}[H]
	\centering
	\begin{tabular}{|c|cc|cc|c| }
		\hline $a_2$ & $Direct(n<3/4)$   &~& $Lensed$ $Ring(3/4<n<5/4)$  &~& $Photon$ $Ring(n>5/4)$ \\
		\hline
		-1.5 & $b<3.43348$ $\|$ $b>4.14597$ &~&  $3.42248<b<3.48281$ $\|$ $3.49571<b<4.14597$ &~& $3.48281<b<3.49571$\\
		\hline
		2 &  $b<5.85459$ $\|$ $b> 7.16084$ &~&  $5.85459<b<6.09435$ $\|$ $6.14689<b<7.16084$ &~&$6.09435<b<6.14689$\\
		\hline
		4 &$b<6.46608$ $\|$ $b> 7.86273$&~&$ 6.46608<b<6.75393$ $\|$ $6.81325<b<7.86273$ &~&$6.75393<b<6.81325$\\
		\hline
		6 &$b<6.96931$ $\|$ $b> 8.42955$&~&$ 6.96931<b<7.29524$ $\|$ $7.35871<b<8.42955$ &~&$7.29524<b<7.35871$\\
		\hline
	\end{tabular}\\
	\caption{The ranges of the impact parameter $b$ corresponding to the direct emission, lensed ring emission, and photon ring emission of the regular BH, given $b_2=0$, for various values of $a_2$.
	}  \label{tab_regular_v4}
\end{table}

We employ the ray-tracing method to solve the trajectory equation \eqref{rphi} and depict the number of photon orbits $n$ as a function of the impact parameter $b$ for various values of $a_2$ and $b_2$ in Fig.\ref{totaln}. In this figure, the red, yellow and black curves represent the physical images corresponding to the photon ring, the lensed ring and the direct emissions, respectively. To provide further details, the ranges of the impact parameter $b$ for the direct, the lensed ring and the photon ring emissions with different values of $a_2$ and $b_2$, are listed in Tables \ref{tab_regular_v2}, \ref{tab_regular_v4}, and \ref{tab_regular_v6}. It is observed that when the parameter $b_2$ is large, the ranges of the impact parameter $b$ corresponding to the lensed ring and the photon ring emissions decrease. Conversely, increasing the parameter $a_2$ enhances the ranges of the impact parameter $b$ for both the lensed ring and the photon ring emissions. When both parameters $a_2$ and $b_2$ are activated simultaneously, the ranges of the impact parameter $b$ for the lensed ring and the photon ring emissions are further expanded. These findings suggest that $a_2$ plays a more significant role in influencing the total orbit number $n$.

\begin{table}[H]
	\centering
	\begin{tabular}{|c|cc|cc|c| }
		\hline  & $Direct(n<3/4)$   &~& $Lensed\  Ring(3/4<n<5/4)$  &~& $Photon\ Ring(n>5/4)$ \\
		\hline 
		0 & $b<5.01514$ $\|$ $b>6.16757$ &~&  $5.01514<b<5.18781$ $\|$ $5.22794<b<6.16757$ &~&$5.18781<b<5.22794$\\
		\hline
		2 & $b<5.40645$ $\|$ $b>6.87491$ &~&  $5.40645<b<5.69251$ $\|$ $ 5.76595<b<6.87491$ &~& $5.69251<b< 5.76595$\\
		\hline
		4 & $b<5.70698$ $\|$ $b>7.47058$ &~&  $5.70698<b<6.10819$ $\|$ $6.22016<b<7.47058$ &~& $6.10819<b<6.22016$\\
		\hline
		6 &  $b<5.95064$ $\|$ $b>7.99503$ &~&  $5.95064<b<6.46598$ $\|$ $6.62078<b<7.99503$ &~&$6.46598<b<6.62078$\\
		\hline
	\end{tabular}\\
	\caption{ The ranges of the impact parameter $b$ corresponding to the direct emission, lensed ring emission, and photon ring emission of the regular BH for various values of $a_2=b_2$.
	}  \label{tab_regular_v6}
\end{table}

\subsection{Observed intensities and optical appearances}\label{Image-thin}

For a static observer, the light emitted from the thin accretion disk appears isotropic. The emission intensity profile for such disks is given by:
\begin{eqnarray}
	I_{\nu}^{em}(r)=I(r)\,,
	\label{Iem-V1}
\end{eqnarray}
where $\nu$ denotes the emission frequency in the static observer’s frame.  

Liouvilles theorem points out that $I_{\nu}(r)/\nu^3$ is conserved along the ray, that is, $I_{\nu}^{em}(r)/\nu^3=I_{\nu'}^{obs}(r)/\nu'^3$ \cite{Peng:2020wun}. Therefore, the observed intensity $I_{\nu'}^{obs}(r)$ at any frequency $\nu'$ is given by:
\begin{eqnarray}
	I_{\nu'}^{obs}(r)=g^3 I_{\nu}^{em}(r)=F(r)^{3/2}I_{\nu}^{em}(r)\,,~~~g=\sqrt{F(r)}\,,
	\label{IOBS-V1}
\end{eqnarray}
where $g$ is the redshift factor.
The total observed intensity can be derived by integrating Eq.\eqref{IOBS-V1}:
\begin{eqnarray}
	I^{obs}(r)=\int I_{\nu'}^{obs}(r)d\nu'=F(r)^{2}I^{em}(r)\,,
	\label{IOBS-V2}
\end{eqnarray}
where $I^{em}$ is the total emitted intensity from the thin accretion disk, usually defined as $I^{em}=\int I_{\nu}^{em} d\nu$ \cite{Gralla:2019xty,Wang:2023vcv}. Considering that the energy is extracted as light passes through the thin disk, the intensity received by the observer should be equal to the sum of the luminosities at all intersection points, which can be expressed as follows:
\begin{eqnarray}
	I_{obs}(b)=\sum_{i} F(r)^{2}I^{em}(r)|_{r=r_i(b)}\,,
	\label{IOBS-V2}
\end{eqnarray}
where $r_i(b)$ is transfer function representing the radial positions at which light with impact parameter $b$ intersects the accretion disk for the $i$-th time, with $i=1,2,3\cdots$.

\begin{figure}[ht]
	\centering
	\includegraphics[width=5.2cm]{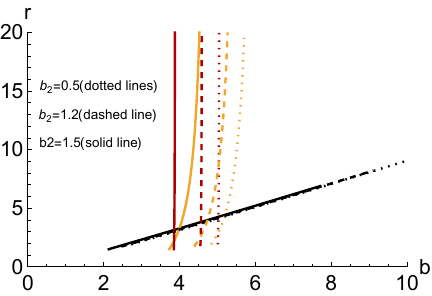}
	\includegraphics[width=5.2cm]{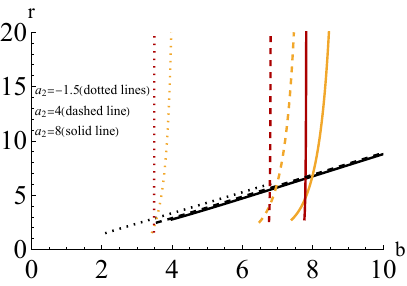}
	\includegraphics[width=5.2cm]{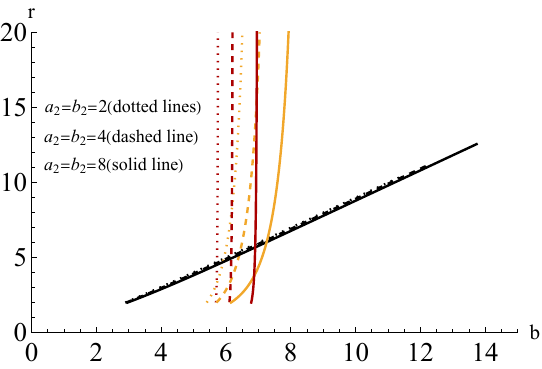}
	\caption{The transfer functions for various values of $a_2$ and $b_2$ are depicted here. The black, yellow and red curves represent the transfer function $r_i(b)$ ($i$=1, 2, 3), respectively. The left plot corresponds to $a_2=0$, the middle plot to $b_2=0$, and the right plot to various values of $a_2=b_2$.}
	\label{figIENRB1}
\end{figure}

The transfer functions $r_i(b)$ as a function of the impact parameter $b$ for various values of the parameters $a_2$ and $b_2$ are illustrated in Fig.\ref{figIENRB1}. They share some common properties: 
\begin{itemize}
    \item The transfer function $r_1(b)$, represented by the black curve, has a slope close to $1$, indicating that it corresponds to the direct image of the redshifted source, which experiences minimal demagnification. In contrast, $r_2(b)$, represented by the yellow curve, has a significantly larger slope, suggesting that its contribution to the total photon flux is relatively small. The transfer function $r_3(b)$, represented by the red curve, exhibits the maximum slope and contributes the least to the total flux.
    \item As illustrated in the first and second plots of Fig.\ref{figIENRB1}, when $a_2=0$ and $b_2$ decreases, the three transfer function curves $r_i$ shift toward larger impact parameters $b$. Conversely, an increase in the parameter $a_2$ has the opposite effect, causing the transfer function curves to move toward smaller impact parameters $b$.
    \item When both parameters $a_2$ and $b_2$ are activated, the impact of the parameter $a_2$ dominates the behavior of the system. This characteristic is particularly noticeable in the image of the regular BH, as will be further elaborated in the following discussion.
\end{itemize}

To visualize and quantitatively study the effects of quantum parameters on the image of the regular BH, we will utilize toy models to describe the radiation emitted by the thin accretion disk at different starting positions. The first model considers radiation emitted from a thin accretion disk located at the innermost stable circular orbit (ISCO). The intensity profile is defined as follows
\begin{eqnarray}
	\text{Profile I}: I_{\text{em}}(r)=\begin{cases}
		I_0[\frac{1}{r-(r_{ISCO}-1)}]^2	& \text{ if } r>r_{ISCO} \\
		0	& \text{ if } r\leq r_{ISCO} 
	\end{cases}\,,
	\label{Profile-V1}
\end{eqnarray}
where $I_0$ is the maximum intensity, and $r_{ISCO}$ is given by
\begin{eqnarray}
	&&
	r_{ISCO}=\frac{3F(r_{ISCO})F'(r_{ISCO})}{2F'(r_{ISCO})^2-F(r_{ISCO})F''(r_{ISCO})}\,.
	\label{risco}
\end{eqnarray}
The second model we consider assumes that the emission starts from the event horizon of the BH. The radiation intensity in this case is described by
\begin{eqnarray}
\text{Profile II}: I_{\text{em}}(r)=\begin{cases}
		I_0[\frac{\frac{\pi}{2}-arctan[r-(r_{ISCO}-1)]}{\frac{\pi}{2}-arctan[r_h-(r_{ISCO}-1)]}	& \text{ if } r>r_{h} \\
		0	& \text{ if } r\leq r_{h} 
	\end{cases}\,.
	\label{Profile-V3}
\end{eqnarray}
Another important accretion disk model assumes that the radiation starts from the photon sphere $r_{ph}$. The image of this IERBH illuminated by such an accretion disk has been previously investigated in Ref. \cite{Cao:2023par}. For the sake of completeness, the main results of the image is also presented in Appendix \ref{appendix-C}.

\begin{figure}[htbp]
	\centering
	\subfigure[\, $a_2=0$]
	{\includegraphics[width=4.2cm]{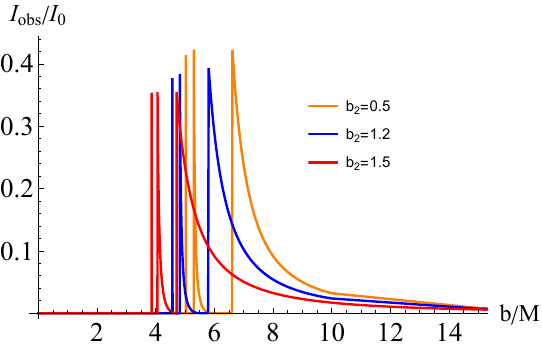} \label{} \hspace{2mm}\includegraphics[width=4cm]{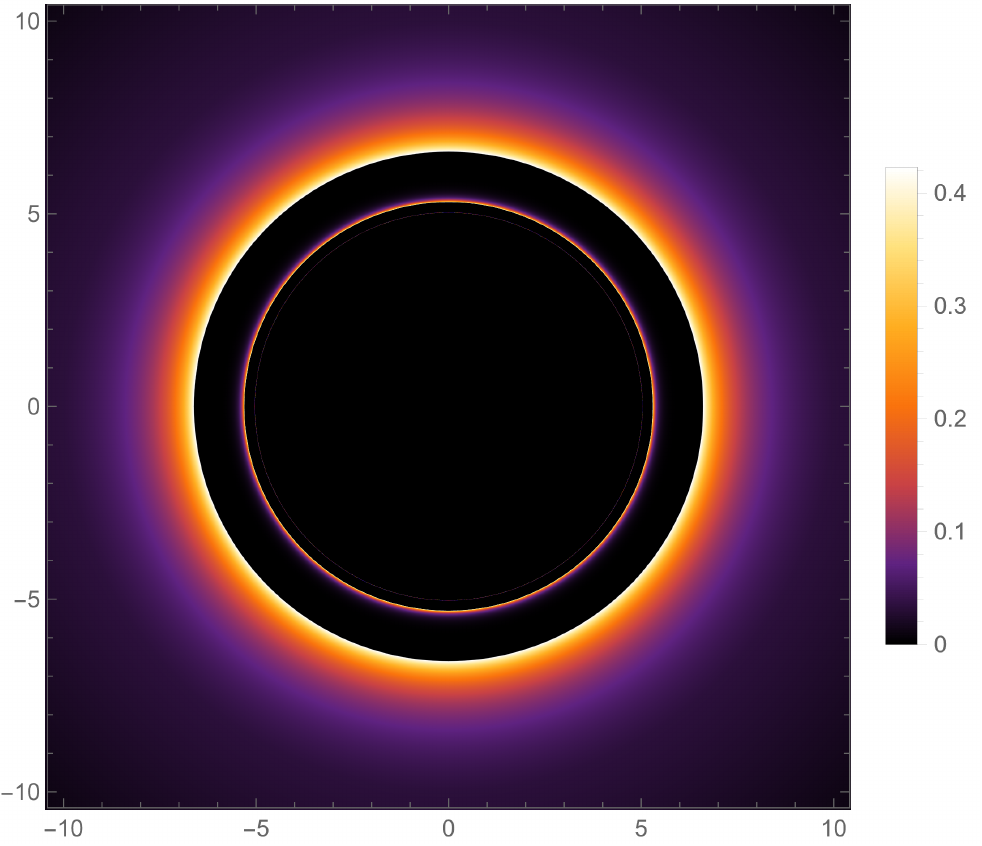} \hspace{2mm} \includegraphics[width=4cm]{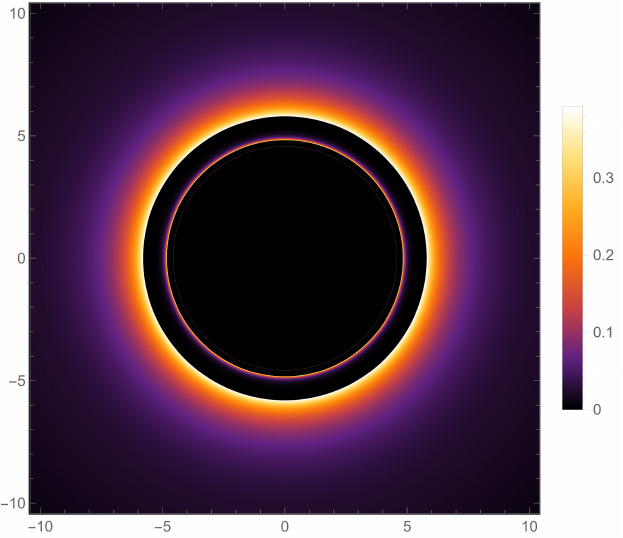}
	\includegraphics[width=4cm]{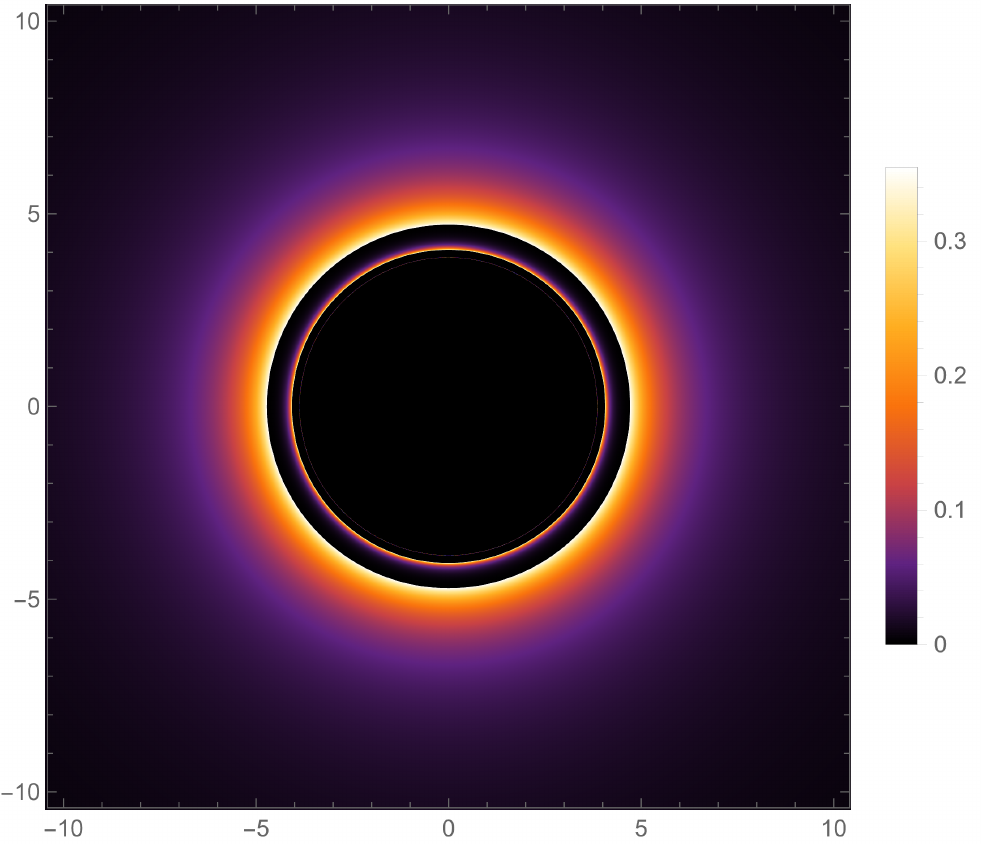}}\\
	\subfigure[\, $b_2=0$]
	{\includegraphics[width=4.2cm]{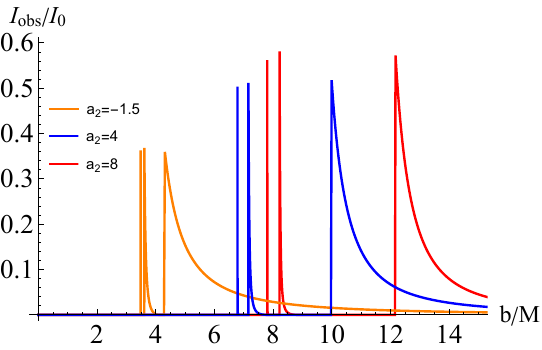} \label{}\hspace{2mm} \includegraphics[width=4cm]{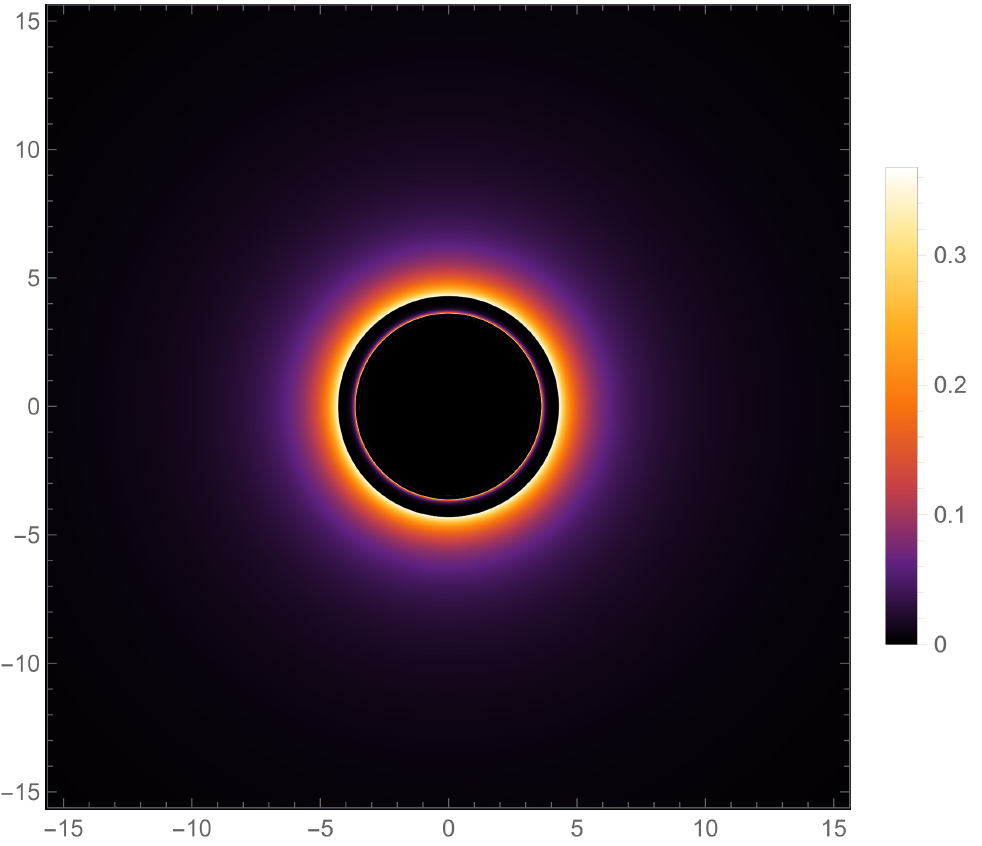}\hspace{2mm} \includegraphics[width=4cm]{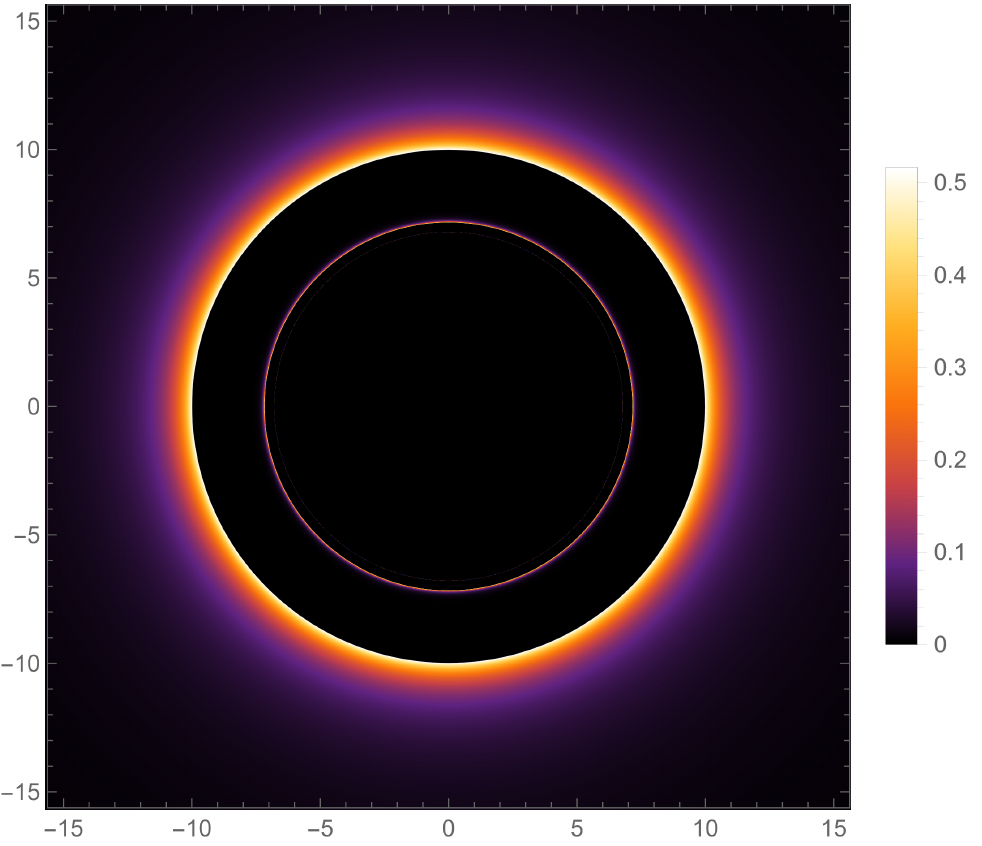}
		\includegraphics[width=4cm]{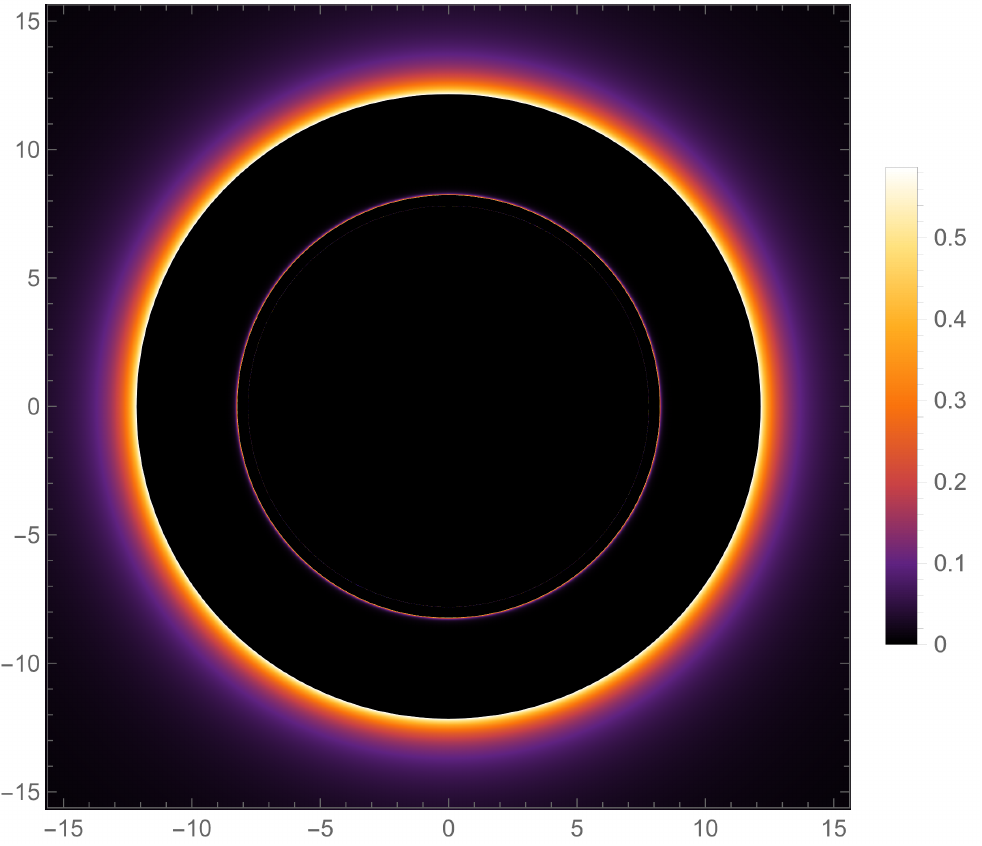}}\\
	\subfigure[\, $a_2=b_2$]
	{\includegraphics[width=4.2cm]{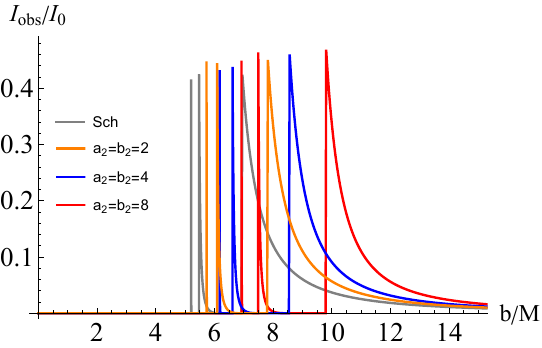} \label{}\hspace{2mm} \includegraphics[width=4cm]{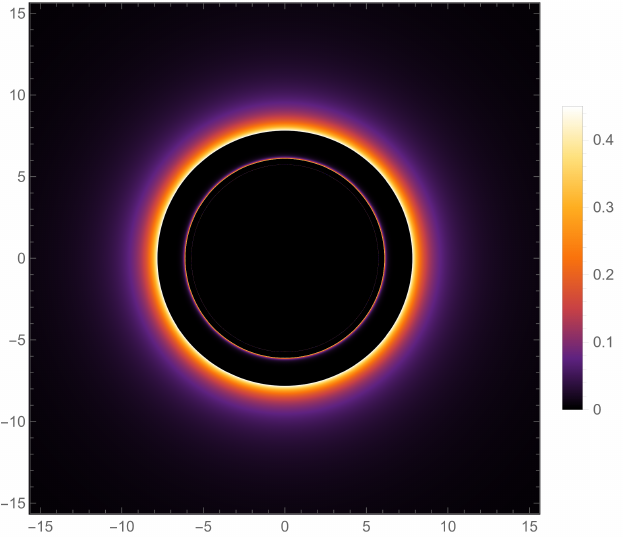}\hspace{2mm} \includegraphics[width=4cm]{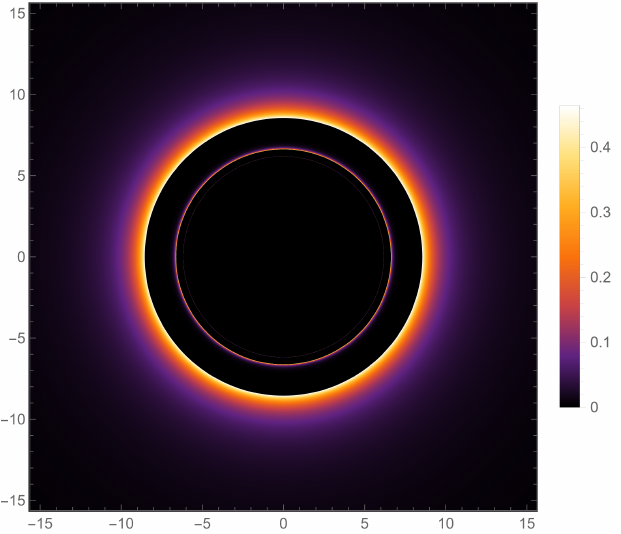}
		\includegraphics[width=4cm]{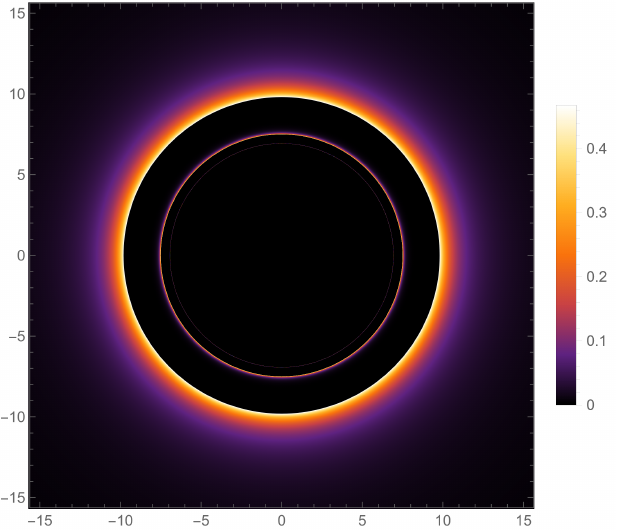}}\\
	\caption{ Observational appearances of the thin accretion disk with intensity profile I in \eqref{Profile-V1}, for various values of $a_2$, $b_2$. The first column shows the total observed intensities $I_{obs}/I_0$ as a function of the impact parameter $b$. The second-fourth column displays the BH optical appearance: the distribution of observed intensities into two-dimensional disks. For $a_2=0$: corresponding to $b_2=0.5,1.2,1.5$. For $b_2=0$: corresponding to $a_2=-1.5,4,8$. For $a_2=b_2$: corresponding to $2,4,8$.}
	\label{figprofile1}
\end{figure}
\begin{figure}[htbp]
	\centering
	\subfigure[\, $a_2=0$]
	{\includegraphics[width=4.5cm]{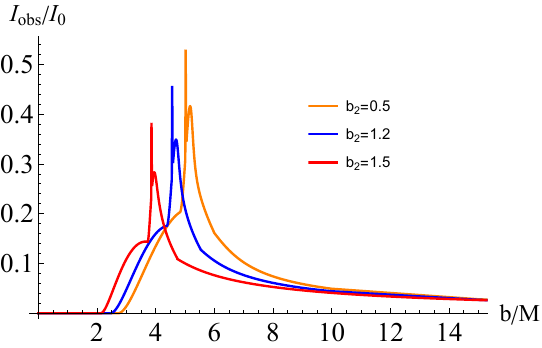} \label{}\hspace{2mm} \includegraphics[width=3.8cm]{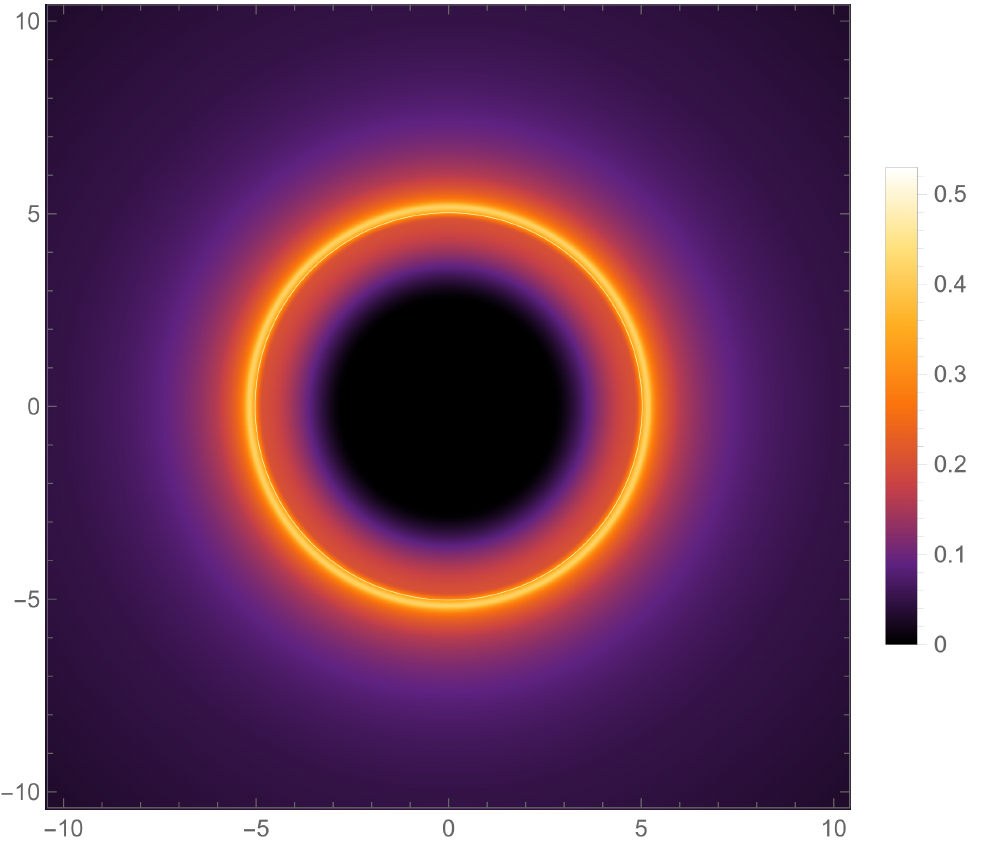}\hspace{2mm} \includegraphics[width=3.8cm]{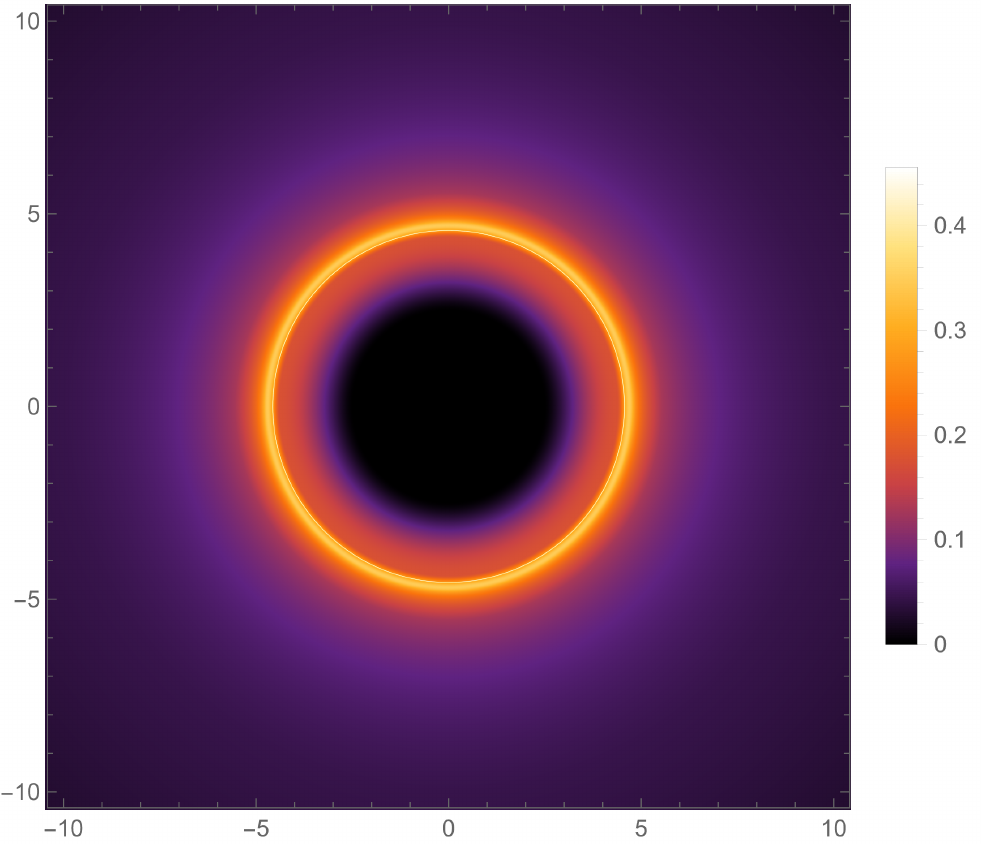}
		\includegraphics[width=3.8cm]{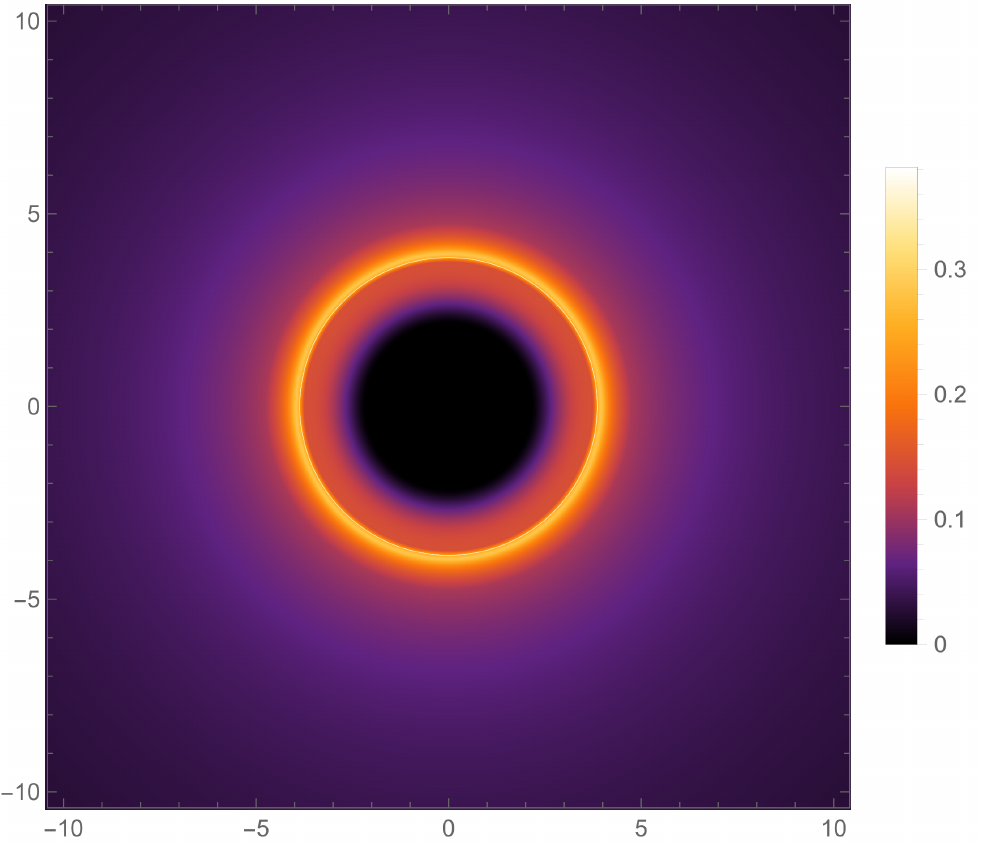}}\\
	\subfigure[\, $b_2=0$]
	{\includegraphics[width=4.5cm]{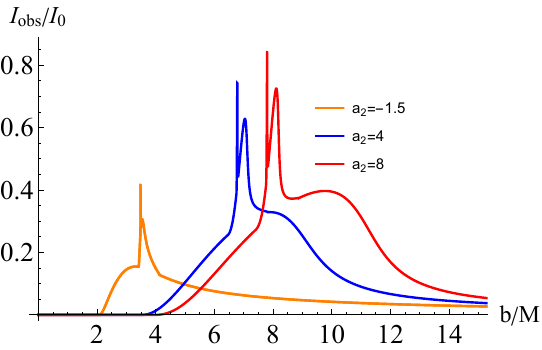} \label{}\hspace{2mm} \includegraphics[width=3.8cm]{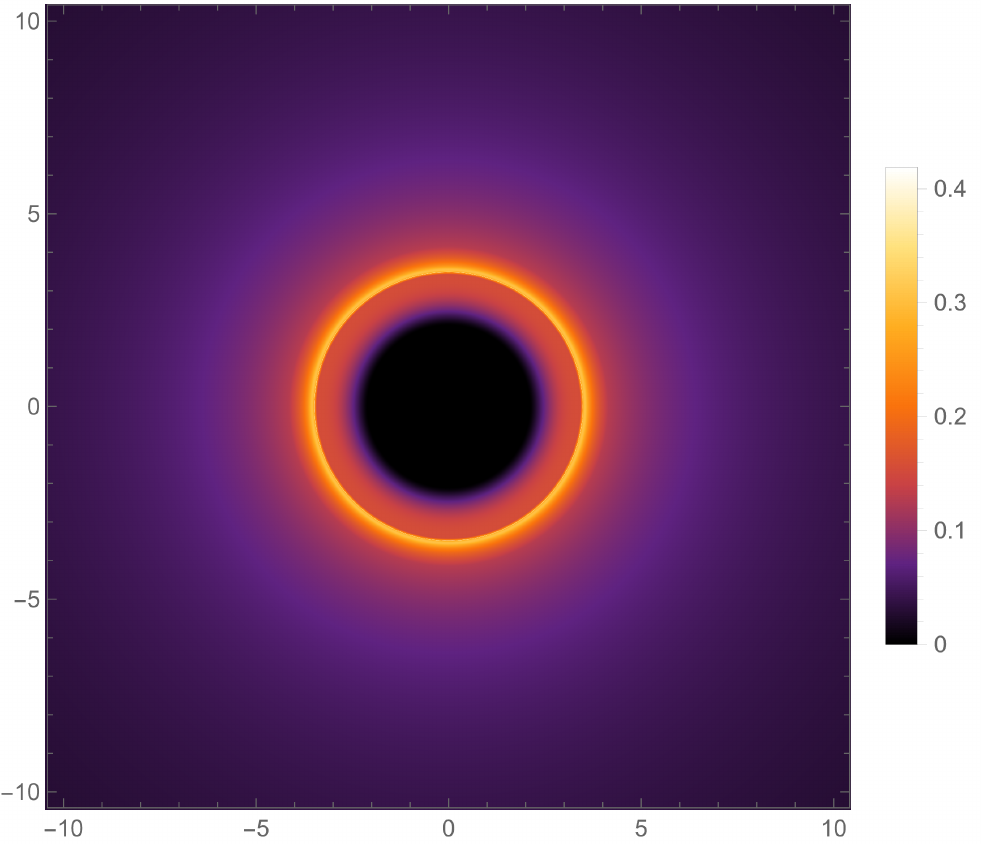}\hspace{2mm} \includegraphics[width=3.8cm]{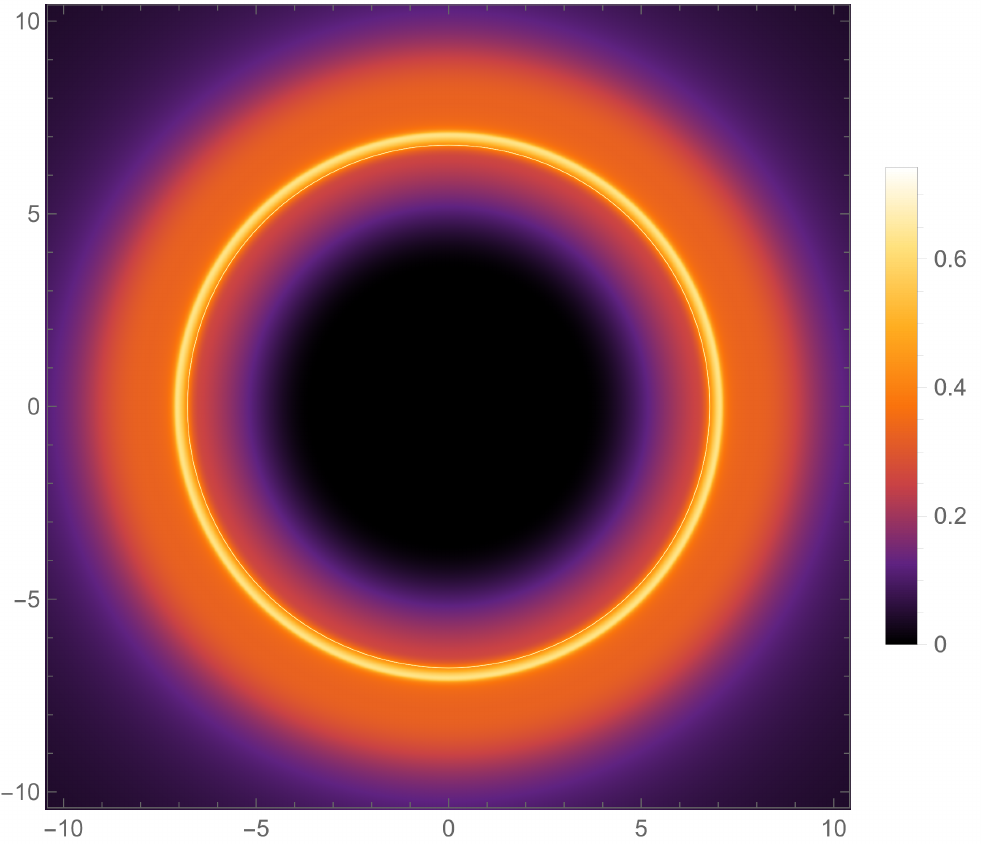}
		\includegraphics[width=3.8cm]{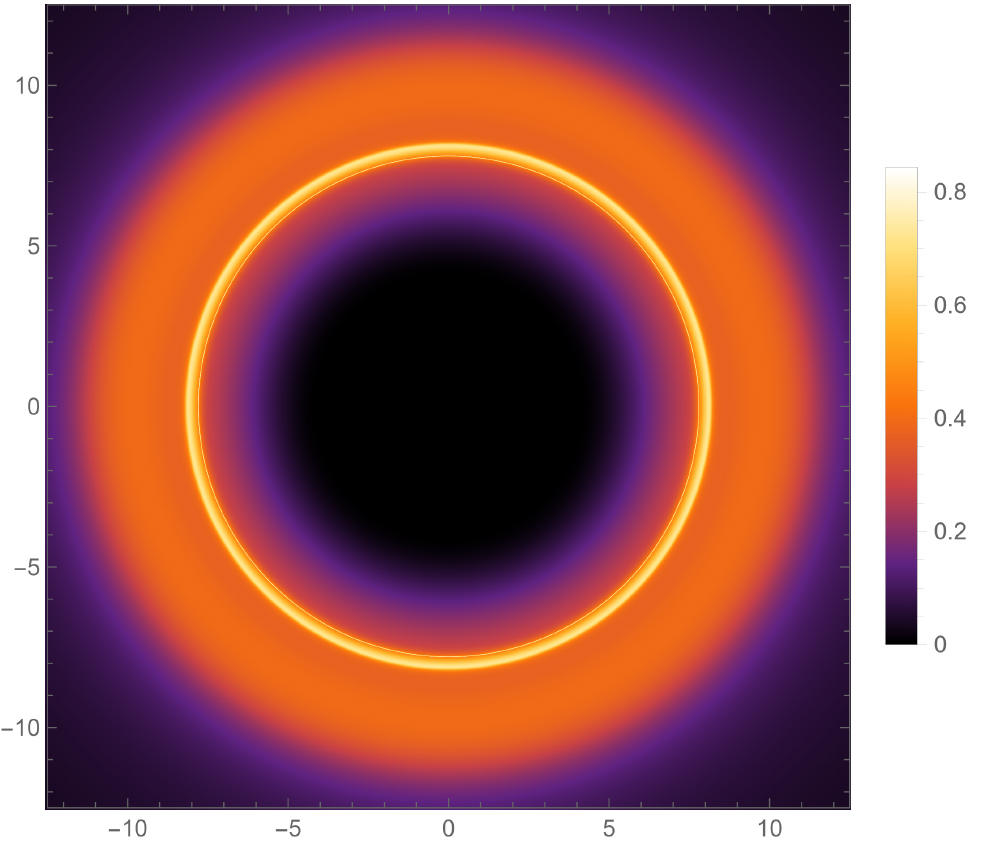}}\\
	\subfigure[\, $a_2=b_2$]
	{\includegraphics[width=4.5cm]{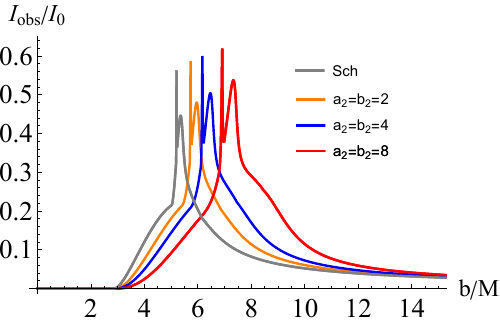} \label{}\hspace{2mm} \includegraphics[width=3.8cm]{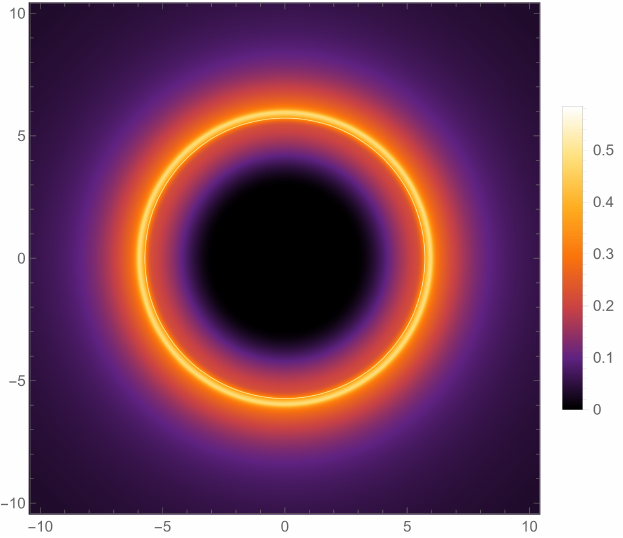}\hspace{2mm} \includegraphics[width=3.8cm]{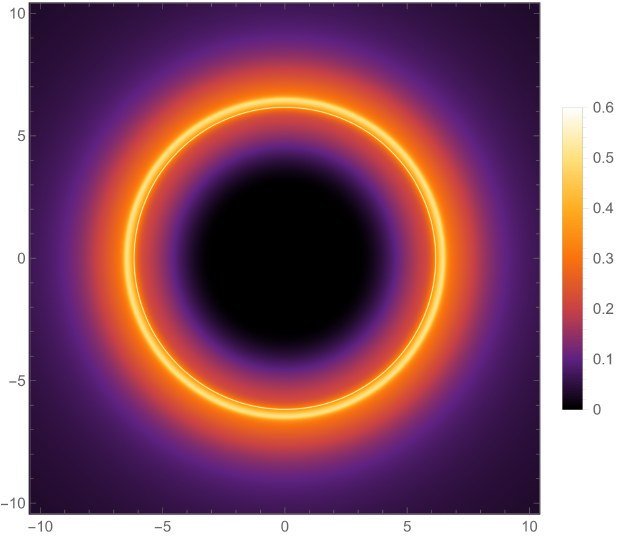}
		\includegraphics[width=3.8cm]{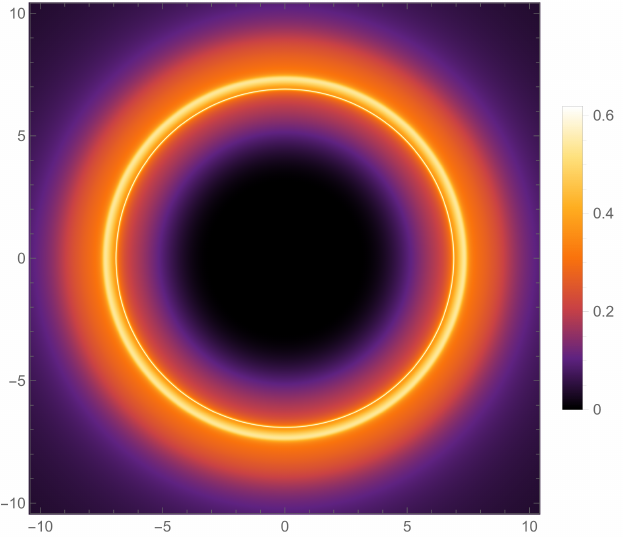}}\\
	\caption{Observational appearances of the thin accretion disk with intensity profile II in \eqref{Profile-V3}, for various values of $a_2$, $b_2$. The first column shows the total observed intensities $I_{obs}/I_0$ as a function of the impact parameter $b$. The second-fourth column displays the BH optical appearance: the distribution of observed intensities into two-dimensional disks. For $a_2=0$: corresponding to $b_2=0.5,1.2,1.5$. For $b_2=0$: corresponding to $a_2=-1.5,4,8$. For $a_2=b_2$: corresponding to $2,4,8$.}
	\label{figprofile3}
\end{figure}

Fig.\ref{figprofile1} presents the observed intensities and images of the regular BH illuminated by the thin accretion disk with profile I. The first figure in each row displays the total observed intensities, $I_{obs}$. The total observed intensities exhibit three distinct peaks, primarily influenced by the photon ring intensity, lensed ring intensity and direct intensity. For large values of $b_2$ with $a_2$ fixed at zero, the observed intensities shift to smaller $b$, resulting in smaller light ring images, as shown in the right plots for the first row in Fig.\ref{figprofile1}. Notably, as the parameter $b_2$ approaches its upper bound, the total observed intensity emitted by this IERBH, characterized by the accretion disk I with the intensity profile in Eq.\eqref{Profile-V1}, shows a significant deviation from that of the Schwarzschild BH. This deviation is highlighted in Fig.\ref{figprofile1fulu} in the Appendix, which illustrates the variation of the total observed intensity for different values of $b_2$, which is close to its upper bound. This pronounced variation is primarily due to the overlap of the direct emission and lensed emission, as indicated by the black and yellow curves in Fig.\ref{figprofile1fulu}. In contrast, increasing the parameter $a_2$ with $b_2=0$ results in a larger shadow region, as demonstrated in the second row of Fig.\ref{figprofile1}. The case where $a_2=b_2$ is particularly intriguing. Although the horizon radius of this regular BH matches that of the Schwarzschild BH, the images in the third row of Fig.\ref{figprofile1} clearly show significant variations in image size as both $a_2$ and $b_2$ increase. These results suggest that the accretion disk model with Profile I can distinctly differentiate between the Schwarzschild BH and this regular BH. The differences in the observed intensities and images, particularly the variations in the light ring and shadow size, provide clear evidence of the distinct nature of the regular BH compared to the Schwarzschild BH.

The observational appearances of this regular BH, illuminated by the accretion disk II with the profile given in Eq.\eqref{Profile-V3}, are presented in Fig.\ref{figprofile3}. In contrast to the scenario I described by Eq.\eqref{Profile-V1}, this case exhibits two distinct peaks in the total observed intensity: a sharp main peak and a slightly lower, relatively wider secondary peak. These features are primarily due to the different contributions from direct emission, lensed ring emission, and photon ring emission. For detailed explanations, please refer to the corresponding discussions in the Appendix \ref{appendix-B}. The influence of varying quantum correction parameters on the total observed intensity and the variation trend of the BH image is consistent with the analysis of the first thin accretion disk model. Specifically, the total observed intensity exhibits significant deviations from that of the Schwarzschild BH. This indicates that the unique characteristics of the observational appearances of the regular BH illuminated by the accretion disk II can effectively differentiate this regular BH from the Schwarzschild BH.

\section{Images of IERBH illuminated by spherical accretions}\label{spherical-img}

Typically, when matter in the universe is captured by a BH and possesses a large angular momentum, a disk-shaped accretion flow forms around the BH. Conversely, when the matter has extremely small angular momentum, it flows radially towards the BH, forming a spherically symmetric accretion flow \cite{Yuan:2014gma}. In this case, the specific intensity of the radiation from the accretion flow, as observed by a distant observer located at $ r=\infty $, is expressed as follows \cite{Bambi:2013nla}
\begin{eqnarray}
	I(\nu_o)=\int_\gamma g^3 j_e(\nu_e)dl_{prop}.
	\label{Profile-V4}
\end{eqnarray}
This equation is derived by integrating the specific emissivity along the photon path $ \gamma $. Here, the redshift factor $ g $ is defined as the ratio $ g = \nu_o / \nu_e $, where $ \nu_o $ and $ \nu_e $ are the observed and emitted photon frequencies, respectively. The term $ j_e(\nu_e) $ denotes the emissivity per unit volume in the rest frame of the emitting matter, which typically follows the relationship $ j_e(\nu_e) \propto \delta(\nu_r - \nu_e) / r^2 $, where $ \nu_r $ is the rest-frame frequency of the emitter \cite{Bambi:2013nla}. Additionally, $ dl_{prop} $ represents the infinitesimal proper length along the photon's path. Based on Eq.(\ref{Profile-V4}), the total observed intensity $I_{obs}$ is given as
\begin{eqnarray}
	I_{obs}=\int_{\nu_o}I(\nu_o)d \nu_o =\int_{\nu_e}\int_{\gamma}g^4 j_e(\nu_e)dl_{prop}d \nu_e
	\label{Profile-V42}
\end{eqnarray}
Then we will investigate the effect of a static and infalling spherical accretion flow on the optical image of this regular BH, respectively.

\subsection{Static spherical accretion flow} 

For a static spherical accretion flow, the redshift factor is $g=F(r)^{1/2}$, and the proper length along the photon path is
\begin{eqnarray}
	dl_{prop}=\sqrt{\frac{dr^2}{F(r)}+r^2d\phi^2}=\sqrt{\frac{1}{F(r)}+r^2\left(\frac{d\phi}{dr}\right)^2}dr
	\label{dlprop-static}
\end{eqnarray}
Substituting these expressions into the equation for the total observed intensity, we obtain
\begin{eqnarray}
	I_{obs}=\int_{\gamma}\frac{F(r)^2}{r^2}\sqrt{\frac{1}{F(r)}+r^2\left(\frac{d\phi}{dr}\right)^2}dr
	\label{Profilestatotal}
\end{eqnarray}
Evidently, the total observed intensity in the static spherical accretion model depends on the radial distance $r$ and the integration interval over the photon path. This dependency highlights the significance of the geometric and gravitational properties of the BH in shaping the observed radiation profile, providing a robust framework for differentiating the optical signatures of regular BHs from those of Schwarzschild BHs.

\begin{figure}[htbp]
	\centering
		\subfigure[\, $a_2=0$]
	{\includegraphics[width=4.2cm]{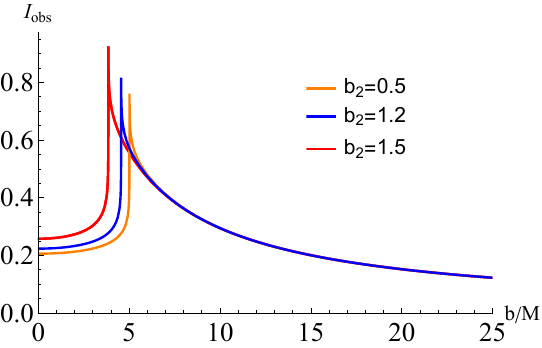} \label{}\hspace{2mm} \includegraphics[width=4cm]{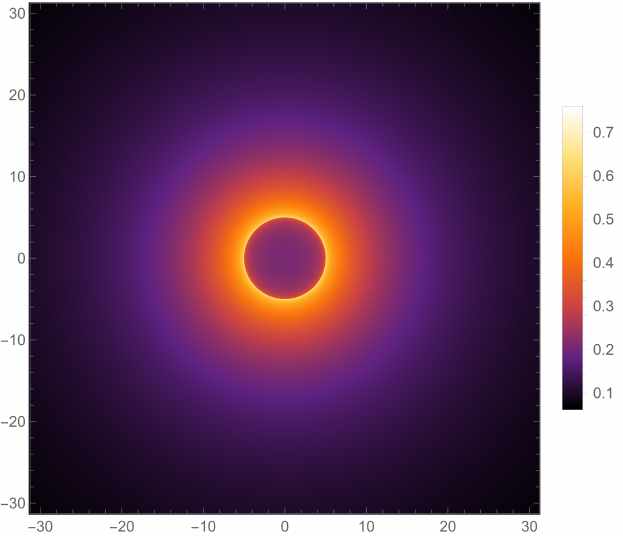}\hspace{2mm} \includegraphics[width=4cm]{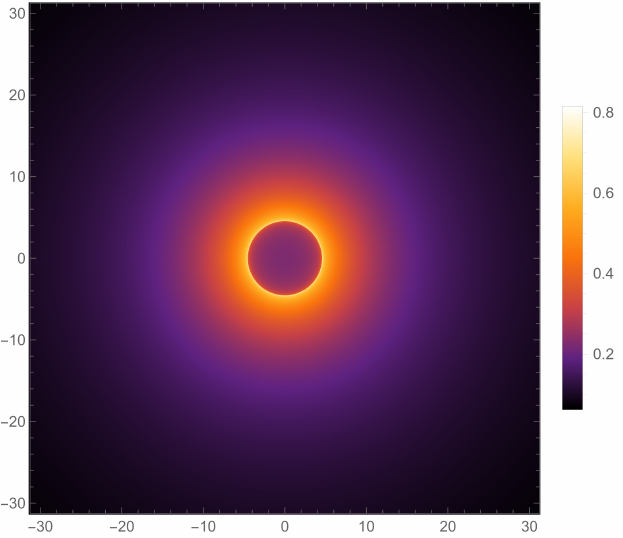}
		\includegraphics[width=4cm]{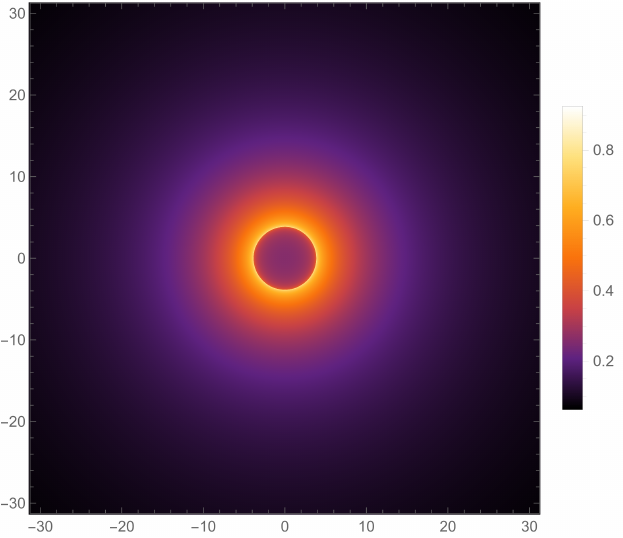}}\\
	\subfigure[\, $b_2=0$]
	{\includegraphics[width=4.2cm]{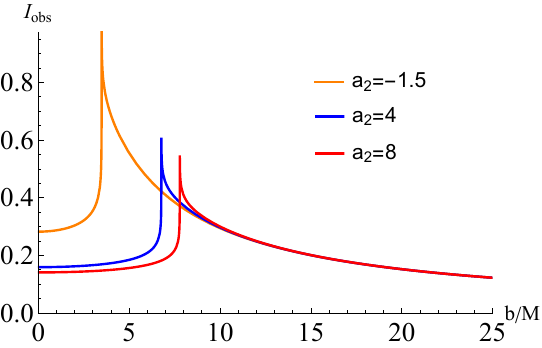} \label{}\hspace{2mm} \includegraphics[width=4cm]{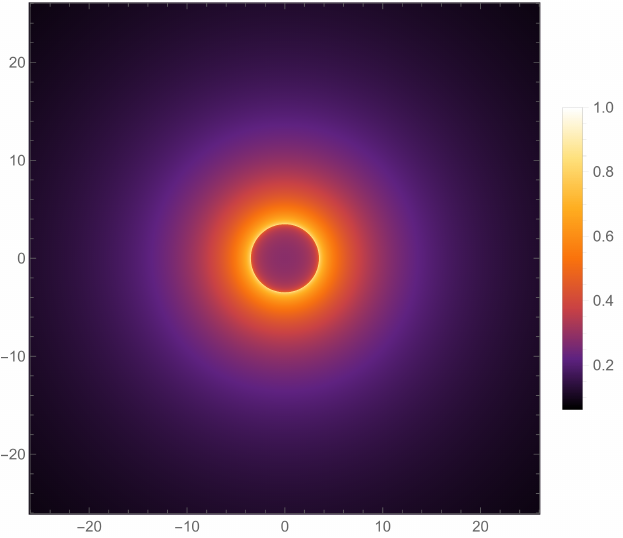}\hspace{2mm} \includegraphics[width=4cm]{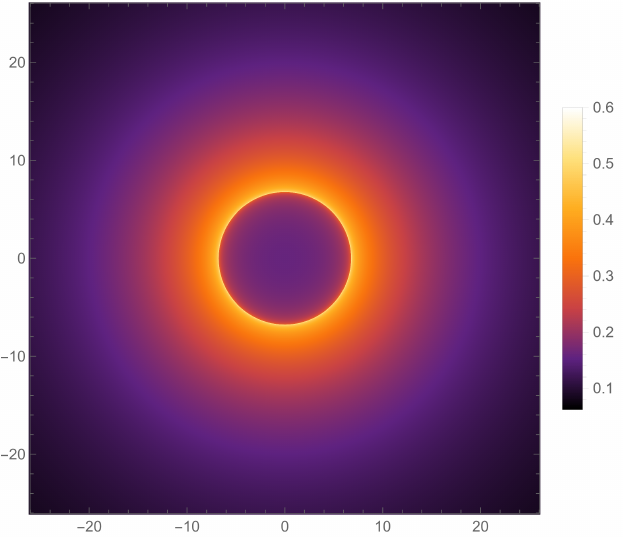}
		\includegraphics[width=4cm]{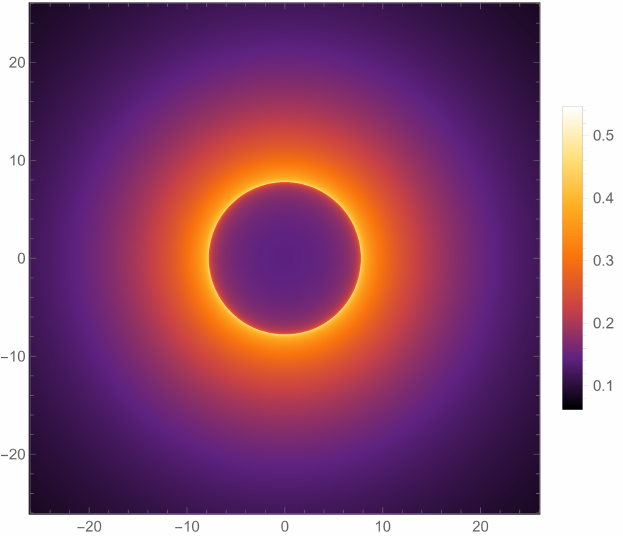}}\\
	\subfigure[\, $a_2=b_2$]
	{\includegraphics[width=4.2cm]{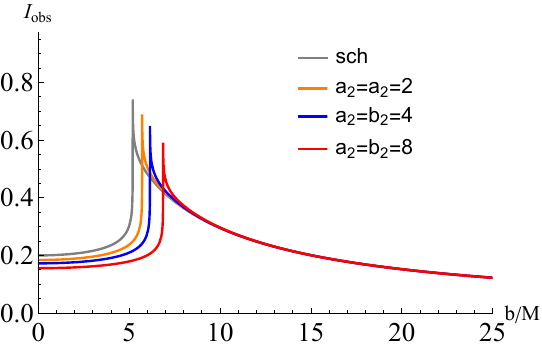} \label{}\hspace{2mm} \includegraphics[width=4cm]{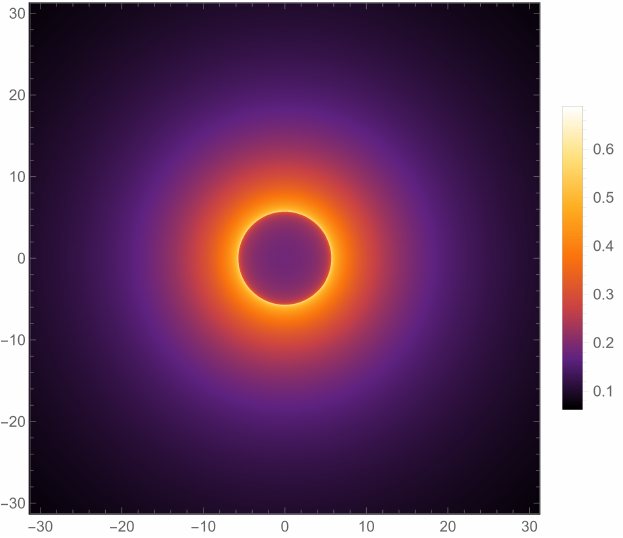}\hspace{2mm} \includegraphics[width=4cm]{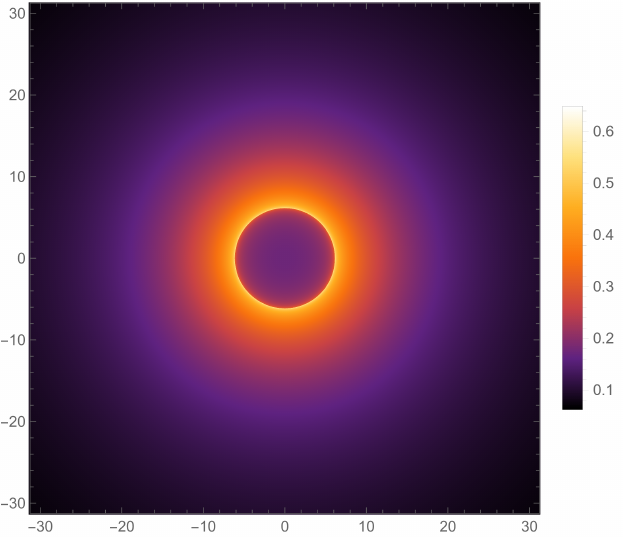}
		\includegraphics[width=4cm]{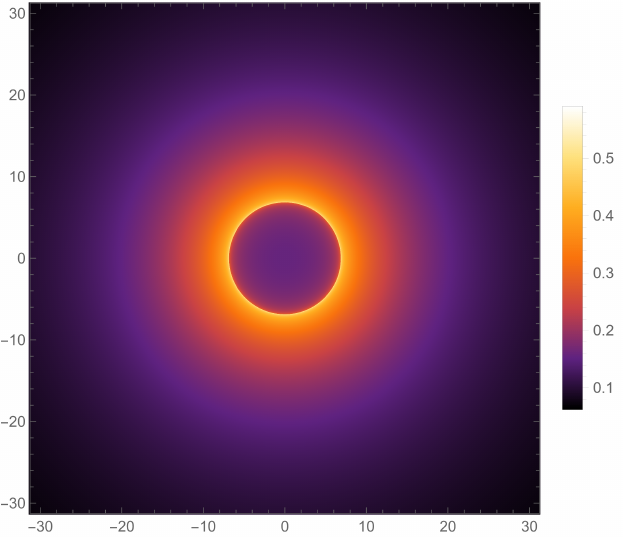}}\\
	\caption{Observational appearances of the static spherical accretion model in \eqref{Profilestatotal}, for various values of $a_2$, $b_2$. The first column shows the total observed intensities $I_{obs}/I_0$ as a function of the impact parameter $b$. The second-fourth column displays the BH optical appearance: the distribution of observed intensities into two-dimensional disks. For $a_2=0$: corresponding to $b_2=0.5,1.2,1.5$. For $b_2=0$: corresponding to $a_2=-1.5,4,8$. For $a_2=b_2$: corresponding to $2,4,8$.
    }
	\label{figprofile4}
\end{figure}

Fig.\ref{figprofile4} displays the total observed intensity and the two-dimensional shadow images of the regular BH illuminated by static spherical accretion. 
From this figure, it is evident that the total observed intensity $I_{obs}$ reaches a maximum at the critical impact parameter $b_{ph}$ and then gradually decreases to lower values as $b$ increases. 
Further, we study how $I_{obs}$ varies with the quantum correction parameters $a_2$ and $b_2$. When $a_2=0$, the total observed intensity $I_{obs}$ increases as the parameter $b_2$ increases, while the critical impact parameter $b_{ph}$ decreases. However, considering only the contribution of $a_2$, $I_{obs}$ exhibits an opposite trend. Specifically, when $a_2$ and $b_2$ act simultaneously, $a_2$ still dominates. This result contrasts with the findings for the regular BH illuminated by the two thin accretion disk models described in Subsection \ref{Image-thin}. Under the influence of the quantum correction parameters, the total observed intensity shows an opposite trend. Additionally, it is clear from the two-dimensional shadow images in Fig.\ref{figprofile4} that different parameters significantly affect the size of the shadow radius and the photon sphere radius. All of this evidence allows us to differentiate regular BHs from classical ones.

\subsection{Infalling spherical accretion flow}

In this subsection, we consider a more realistic scenario by introducing a radially infalling spherical accretion, given that the majority of matter in the universe is dynamical. For simplicity, we still employ Eq.(\ref{Profile-V4}) to analyze the image of the regular BH illuminated by the infalling spherical accretion flow. A key distinction from the static spherical accretion flow model is the redshift factor $g$, which is given by 
\begin{eqnarray}
g=\frac{k_{\mu}u^{\mu}_o}{k_{\nu}u^{\nu}_e}\,,
	\label{infallred}
\end{eqnarray}
where $k_{\mu}=\partial L/\partial \dot{x}^{\mu}$ represents the four-momentum of the photon emitted by the accretion matter. This redshift factor accounts for the relative motion between the emitting and observing points, reflecting the dynamic nature of the accretion flow. Based on Eqs.(\ref{td})-(\ref{rdot}), the four-momentum $k_{\mu}$ takes the following form
\begin{eqnarray}
k_{\mu}=(k_{t},k_r,k_{\theta},k_{\phi})=\left(-\frac{1}{b},\pm\frac{1}{F(r)}\sqrt{\frac{1}{b^2}-\frac{F(r)}{r^2}},0,\pm1\right)\,,
	\label{infallred}
\end{eqnarray}
with the $+(-)$ sign in $k_r$ indicating that the photon is moving towards (away from) the BH. The $\pm$ sign in $k_{\phi}$ designates whether the photon is moving counterclockwise or clockwise. The four-velocity of the distant observer $u^{\mu}_o$ is given by $u^{\mu}_o=(1,0,0,0)$.
Additionally, the four-velocity of the infalling matter $u^{\mu}_e$ is 
\begin{eqnarray}
(u^{t}_e,u^{r}_e,u^{\theta}_e,u^{\phi}_e)=\left(\frac{1}{F(r)},-\sqrt{1-F(r)},0,0\right)\,.
	\label{infallumue}
\end{eqnarray}
In this case, the redshift factor $g$ is
\begin{eqnarray}
	g=\frac{k_{t}}{k_{t}u^{t}_e+k_{r}u^{r}_e}=\frac{1}{\frac{1}{F(r)}\pm \sqrt{1-F(r)}\sqrt{\frac{1}{F(r)}(\frac{1}{F(r)}-\frac{b^2}{r^2})}}\,.
	\label{infallg}
\end{eqnarray}
At the same time, the proper length $dl_{\text{prop}}$ is given by \cite{Bambi:2013nla}
\begin{eqnarray}
	dl_{prop}=k_{\mu}u^{\mu}_ed\lambda=\frac{k_t}{g\rvert k_r\rvert}dr\,.
	\label{dlprop-infall}
\end{eqnarray}
Finally, the total observed intensity of the regular BH illuminated by the radially infalling accretion flow is
\begin{eqnarray}
I_{obs}=\int_{\gamma}\frac{g^3k_t}{r^2\rvert k_r\rvert}dr\,.
	\label{inten-infall}
\end{eqnarray}
\begin{figure}[htbp]
	\centering
	\subfigure[\, $a_2=0$]
	{\includegraphics[width=4.2cm]{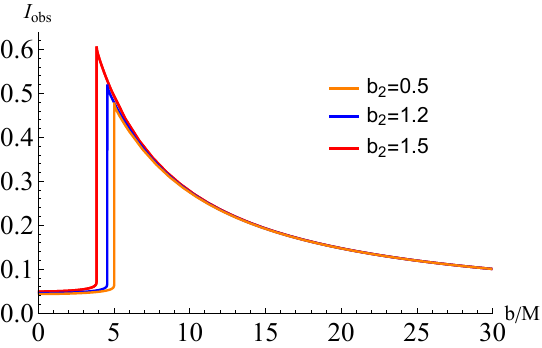} \label{}\hspace{2mm} \includegraphics[width=4cm]{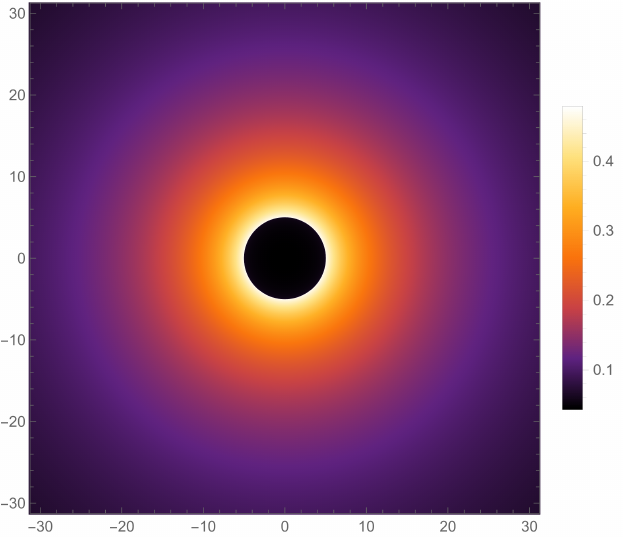}\hspace{2mm} \includegraphics[width=4cm]{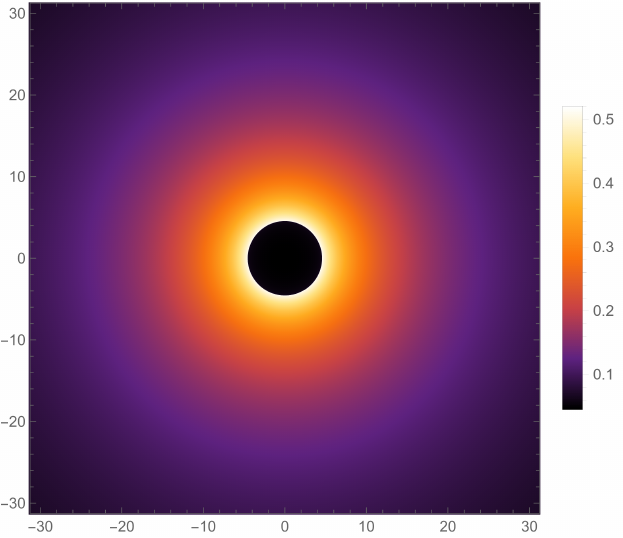}
		\includegraphics[width=4cm]{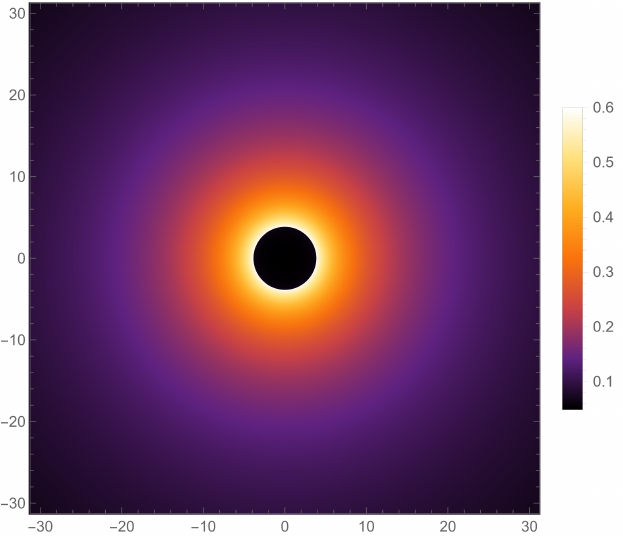}}\\
	\subfigure[\, $b_2=0$]
	{\includegraphics[width=4.2cm]{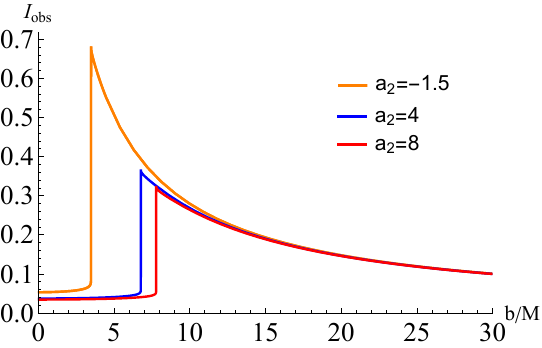} \label{}\hspace{2mm} \includegraphics[width=4cm]{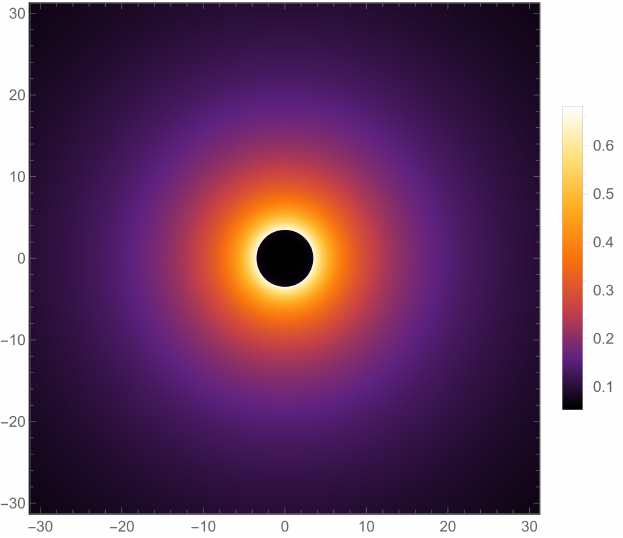}\hspace{2mm} \includegraphics[width=4cm]{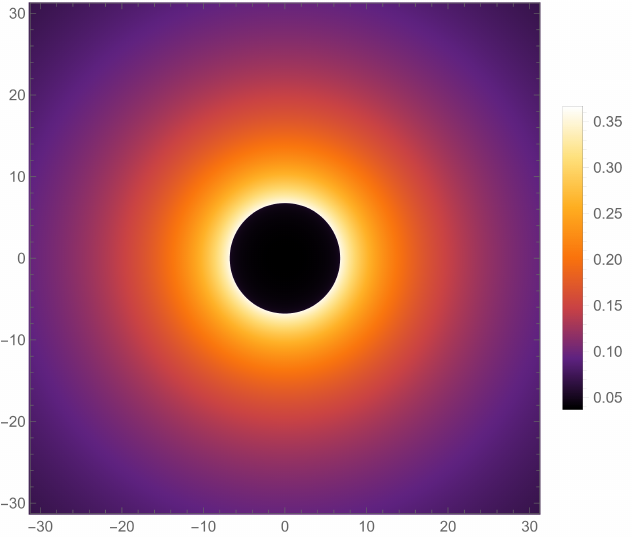}
		\includegraphics[width=4cm]{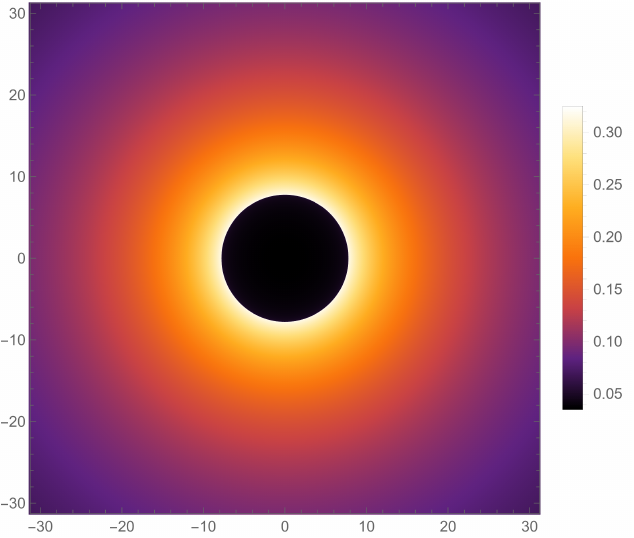}}\\
	\subfigure[\, $a_2=b_2$]
	{\includegraphics[width=4.2cm]{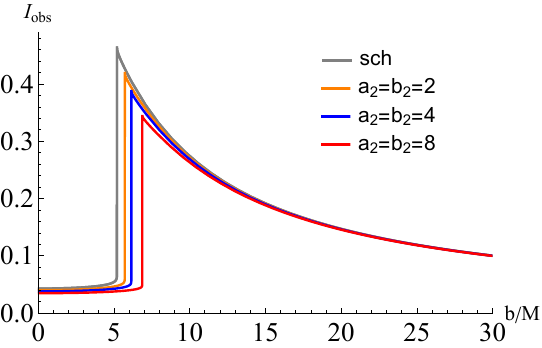} \label{}\hspace{2mm} \includegraphics[width=4cm]{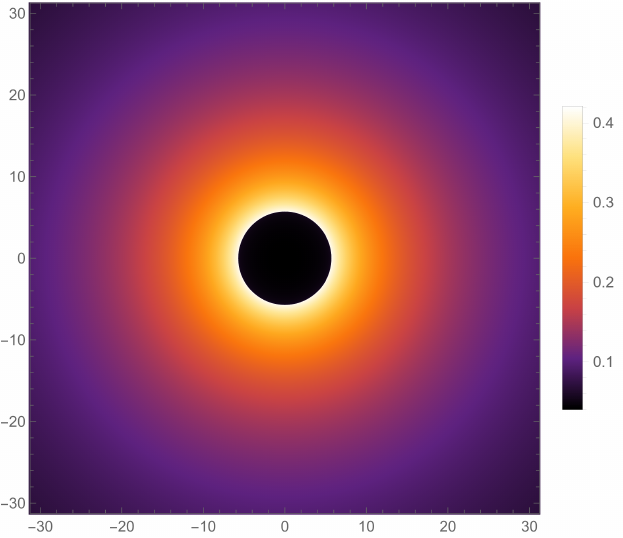}\hspace{2mm} \includegraphics[width=4cm]{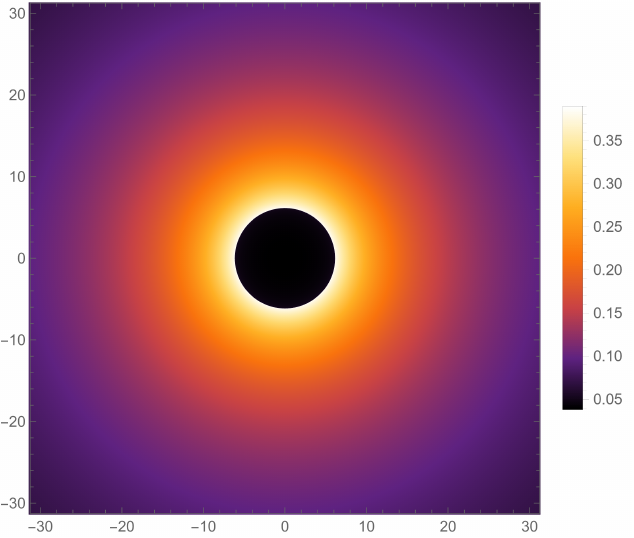}
		\includegraphics[width=4cm]{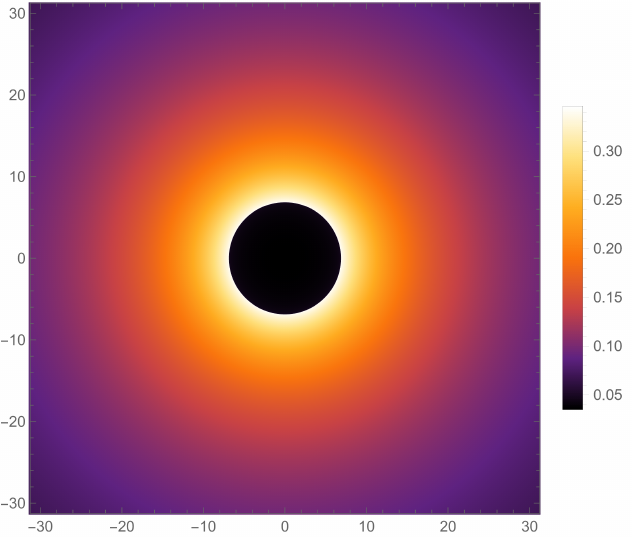}}\\
	\caption{Observational appearances of the infalling spherical accretion model in \eqref{inten-infall}, for various values of $a_2$, $b_2$. The first column shows the total observed intensities $I_{obs}/I_0$ as a function of the impact parameter $b$. The second-fourth column displays the BH optical appearance: the distribution of observed intensities into two-dimensional disks. For $a_2=0$: corresponding to $b_2=0.5,1.2,1.5$. For $b_2=0$: corresponding to $a_2=-1.5,4,8$. For $a_2=b_2$: corresponding to $2,4,8$.}
	\label{figprofile5}
\end{figure}

The first column in Fig.\ref{figprofile5} shows the total observed intensities $I_{obs}/I_0$ as a function of the impact parameter $b$. Similar to the scenario illuminated by static accretion, the total observed intensity reaches a maximum at $b=b_{ph}$ and then gradually decreases as $b$ increases. When compared to the scenario illuminated by the static accretion depicted in Fig.\ref{figprofile4}, we find that the corresponding $b_{ph}$ values are identical for the same quantum correction parameters. However, the total observed intensity $I_{obs}$ in the infalling spherical accretion scenario is consistently lower than that in the static accretion scenario. This difference can be attributed to the Doppler effect, which influences the observed intensity due to the relative motion of the infalling matter. In addition, the quantum correction parameter also has a significant impact on $I_{obs}$, with trends similar to those observed in the static accretion scenario. 

Fig. \ref{figprofile5} also presents the two-dimensional image of the BH shadow and the photon sphere. It is evident that the BH shadow radius and the photon sphere radius in the infalling spherical accretion scenario are the same as those in the static accretion scenario. However, a notable difference can be observed in the images: the infalling spherical accretion scenario results in a darker shadow region in the center compared to the static accretion scenario. This phenomenon, which is more pronounced near the BH event horizon, can be attributed to the Doppler effect. The Doppler effect modulates the observed intensity due to the relative motion of the infalling matter, leading to the observed darkening in the shadow region. Furthermore, the influence of the quantum correction parameter is evident in the changes to the photon ring and the BH shadow radius. These changes distinguish the scenario from that of a Schwarzschild BH. By adjusting the quantum correction parameter, observers can more easily differentiate between various BH types.

\section{ Conclusion and discussion}\label{conslusion}

The image of a BH is influenced by both its geometric properties and the characteristics of the accretion model. In this paper, we have investigated the observational appearances of an IERBH illuminated by different accretion models to examine how these models affect the BH images and to explore the detectability of quantum gravity effects in the optical images of BHs.

While the regular BH is illuminated by the thin accretion disk III, the total observed intensity and the image show only minor differences compared to the Schwarzschild BH scenario \cite{Cao:2023par}, significant variations are observed when the BH is illuminated by thin accretion disks I and II, as well as by a spherical accretion flow. These variations are primarily influenced by the quantum correction parameters $a_2$ and $b_2$, which play a crucial role in affecting the total observed intensities, the BH images, including the shadow region and the photon ring. Specifically, different accretion models highlight distinct features in the BH images. For example, the thin accretion disks I and II produce images where the total observed intensity and the shadow size show clear deviations from the Schwarzschild BH. These deviations are due to the unique contributions from direct emission, lensed ring emission, and photon ring emission. The presence of these features in the images of the regular BH provides a clear signature that can be used to differentiate it from classical BHs.

Moreover, the infalling spherical accretion flow, which is more realistic and accounts for the dynamical nature of matter in the universe, also exhibits significant variations. The total observed intensity and the shadow radius of the regular BH are notably affected by the quantum correction parameters. This dynamical flow results in a darker central region in the BH image, attributed to the Doppler effect, which modulates the observed intensity due to the relative motion of the infalling matter. These findings emphasize the importance of considering the dynamic nature of accretion flows when interpreting BH images. The variations in the light ring and shadow size, influenced by the quantum correction parameters, provide clear evidence of the distinct nature of the regular BH compared to the Schwarzschild BH, thus facilitating the detection of quantum gravity effects.

In summary, by tuning the quantum correction parameters, we can observe distinct changes in the photon ring, shadow radius, and total observed intensities. These changes serve as indirect evidence for quantum gravity effects. On the other hand, when the BH is illuminated by different accretion flows, the quantum gravity effects become more pronounced. These observations are particularly significant because it opens a new window for exploring and verifying theories of quantum gravity in extremely curved spacetimes.

The EHT has provided unprecedented high-resolution images of supermassive BHs in M87* and Sgr A*. Our theoretical predictions can guide the development of observational techniques and data analysis methods, helping to refine and interpret these images with greater precision. The ability to detect subtle differences in BH shadows and photon rings can enhance our understanding of the underlying physics.
The distinctive features in the BH images, such as the darkened central region and the altered shadow radius, can be used to differentiate between regular BH and classical BH. This is particularly important for understanding the nature of compact objects at the centers of galaxies and for testing the predictions of different BH models.

Future research can extend our analysis to include other types of accretion models, such as thick disk accretion \cite{Chen:2024jsv,Witzany:2017zrx,Mewes:2015gma,Font:2002ci}, to provide a more comprehensive understanding of the optical properties of the regular BHs. In addition, combining observations from different wavelengths, such as mm/sub-mm and X-ray bands \cite{Asada:2017hky,Remillard:2006fc,Bambi:2020jpe,Johnson:2024ttr}, can offer additional insights into the nature of regular BHs. This multi-wavelength approach can help verify the predictions of quantum gravity effects and improve the accuracy of BH classification. Detailed numerical simulations of the accretion models and the resulting BH images can further validate our theoretical predictions, refine the models, and provide a more realistic representation of the observed phenomena.

\acknowledgments
	
We are deeply grateful to Yen Chin Ong for the valuable discussions. This work is supported by the Natural Science Foundation of China under Grants Nos. 12447151, 12347159, 12275079, 12035005, 12375055 and 12447137.

\appendix

\section{The observational appearances of thin accretion disk}\label{method}

For the sake of completeness, this appendix provides an overview of the observational appearances of thin accretion disks, and specifically analyzes their observed intensities.
 
\subsection{Thin accretion disk I}\label{appendix-A}

\begin{figure}[htbp]
	\centering
	\subfigure[\, $b_2=5$]
	{\includegraphics[width=4.2cm]{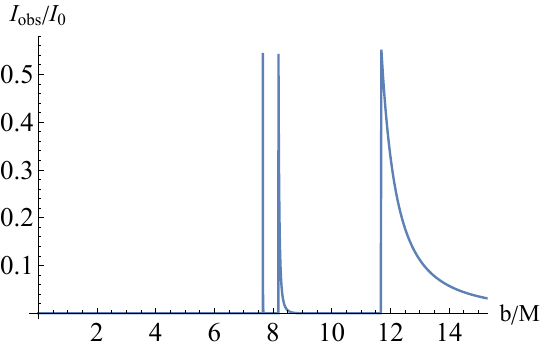} \label{}\hspace{2mm} \includegraphics[width=4.2cm]{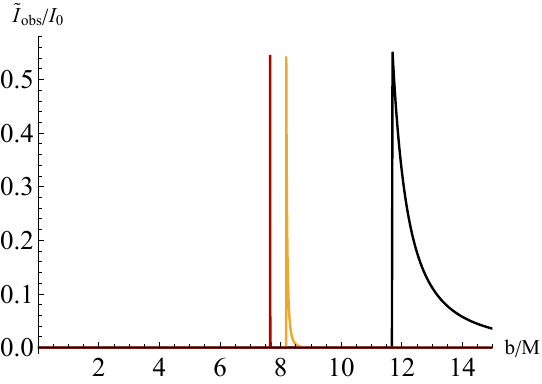}\hspace{2mm} \includegraphics[width=3.6cm]{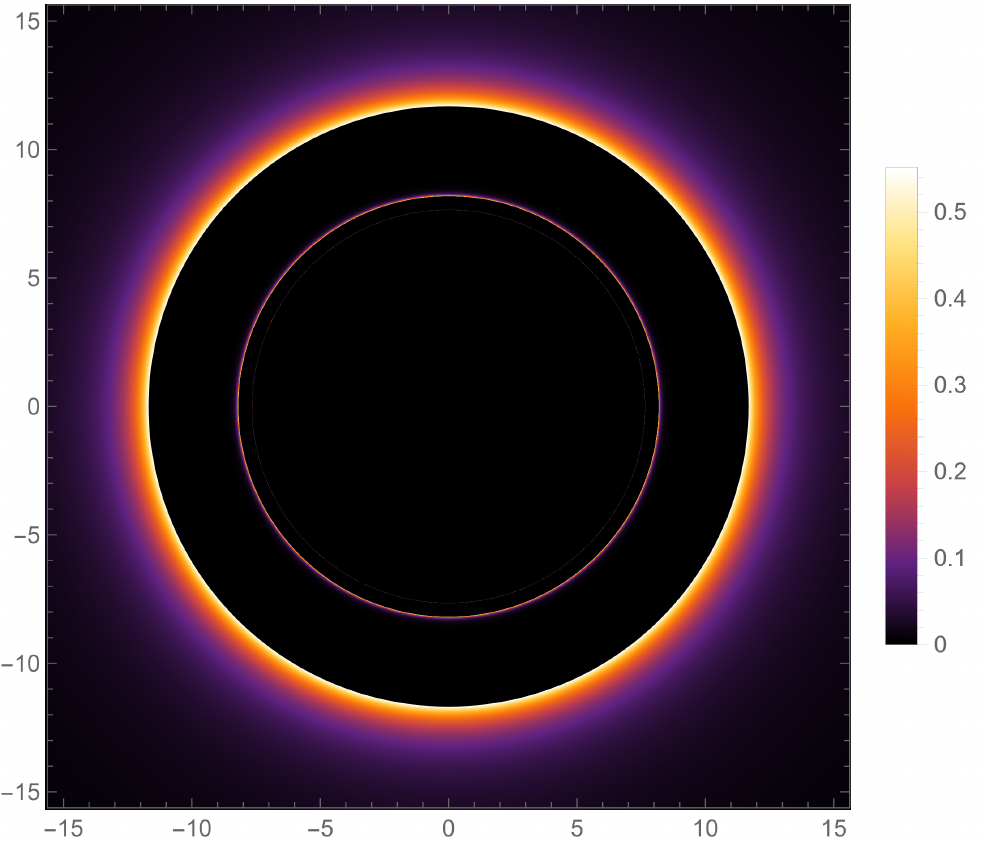}
		\includegraphics[width=3.5cm]{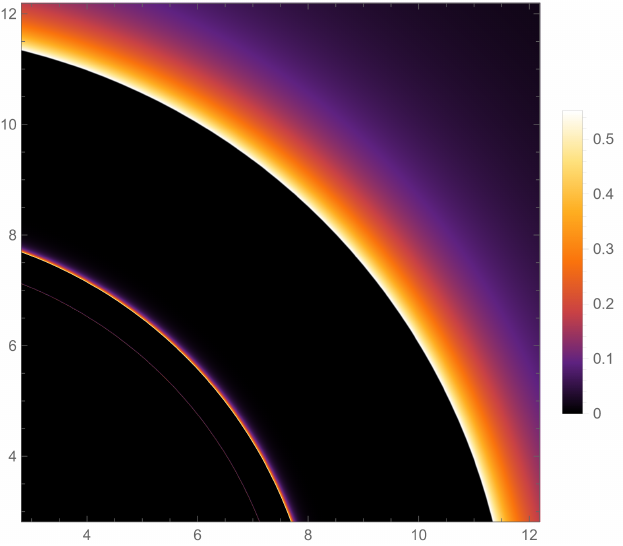}}\\
	\subfigure[\, $b_2=8$]
	{\includegraphics[width=4.2cm]{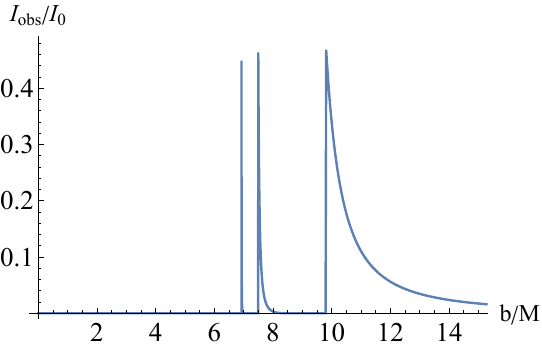} \label{}\hspace{2mm} \includegraphics[width=4.2cm]{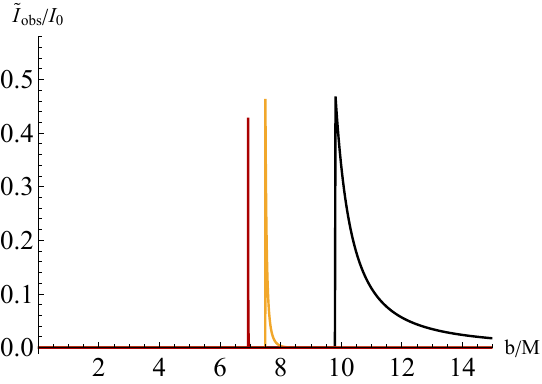}\hspace{2mm} \includegraphics[width=3.6cm]{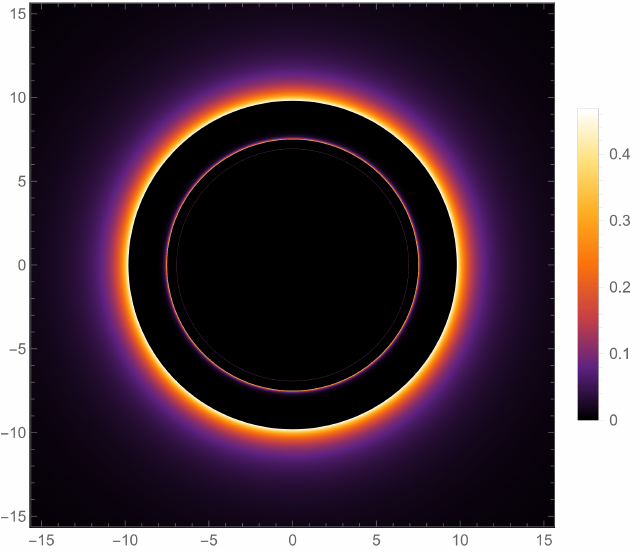}
		\includegraphics[width=3.5cm]{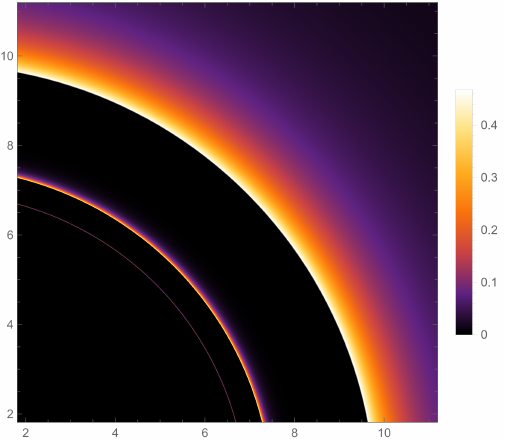}}\\
	\subfigure[\, $b_2=9.5$]
	{\includegraphics[width=4.2cm]{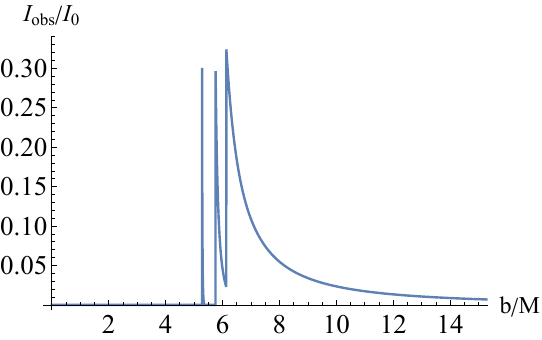} \label{}\hspace{2mm} \includegraphics[width=4.2cm]{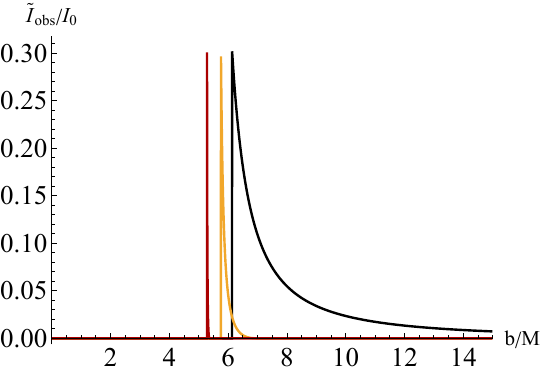}\hspace{2mm} \includegraphics[width=3.6cm]{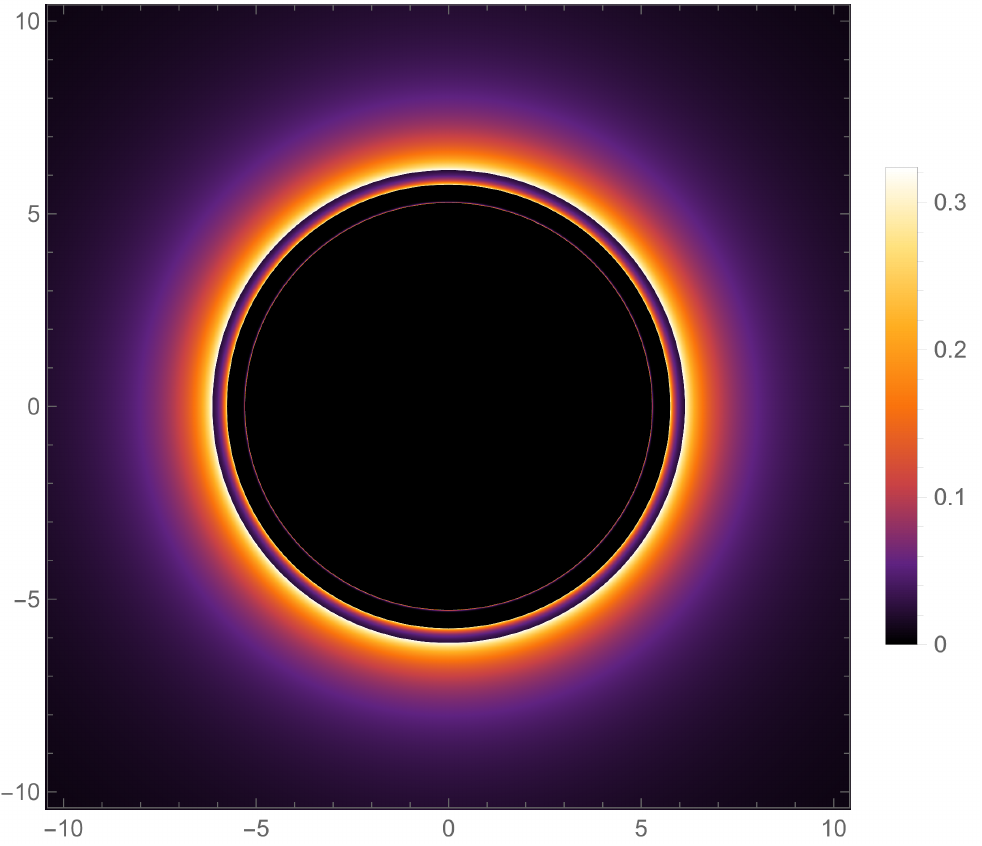}
		\includegraphics[width=3.5cm]{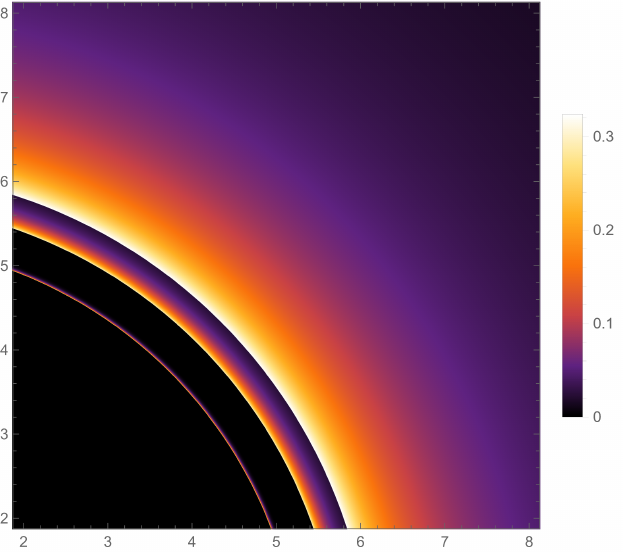}}\\
	\caption{Observational appearances of the thin accretion disk with intensity profile I in \eqref{Profile-V1}, for various values of $b_2$ and $a_2=8$. The first column shows the total observed intensities $I_{obs}/I_0$ as a function of the impact parameter $b$. The second column displays the individual observed intensities corresponding to the first (black), second (yellow), and third (red) transfer functions in Eq.~\eqref{IOBS-V2}. The third column depicts the optical appearances, illustrating the distribution of observed intensities over two-dimensional plane. The fourth column provides zoomed-in views of these optical appearance.}
	\label{figprofile1fulu}
\end{figure}

Fig.\ref{figprofile1fulu} shows the observational appearances of this regular BH, which is illuminated by the thin accretion disk located at ISCO, as described by the profile I of the emitted intensity in Eq.\eqref{Profile-V1}. The first column presents the total observed intensities $I_{obs}/I_0$ as a function of the impact parameter $b$, with a fixed value of $a_2=8$ for various values of $b_2$. The second column displays the individual observed intensities corresponding to the first (black), second (yellow), and third (red) transfer functions in Eq.~\eqref{IOBS-V2}. It is evident that three distinct peaks are presented in the total observed intensities. As $b_2$ increases, all three peaks shift to the left. Notably, the third peak moves the fastest, followed by the second, while the first peak moves the slowest, resulting in a reduced gap between the peaks. Specially, when $b_2$ increases to $b_2=9.5$, the second and third peaks begin to overlap, leading to distinctly different optical features compared to the cases where $b_2=5$ and $b_2=8$. Additionally, the third and fourth columns display the two-dimensional images corresponding to the total observed intensities. Clearly, as $b_2$ increases, the width of the observed intensity distribution also increases, providing a detectable effect in the images.

\subsection{Thin accretion disk II}\label{appendix-B}

\begin{figure}[htbp]
 	\centering
 	\subfigure[\, $b_2=5$]
 	{\includegraphics[width=4.2cm]{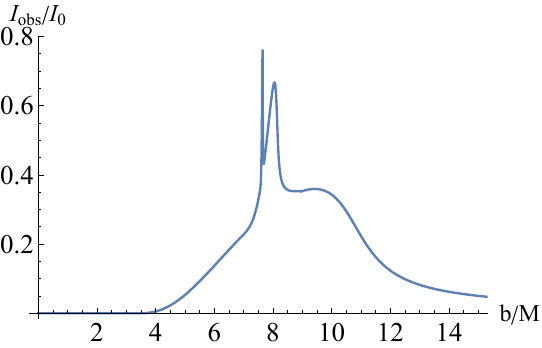} \label{}\hspace{2mm} \includegraphics[width=4.2cm]{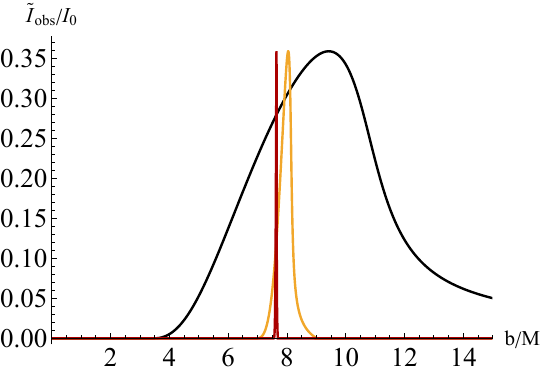}\hspace{2mm} \includegraphics[width=3.6cm]{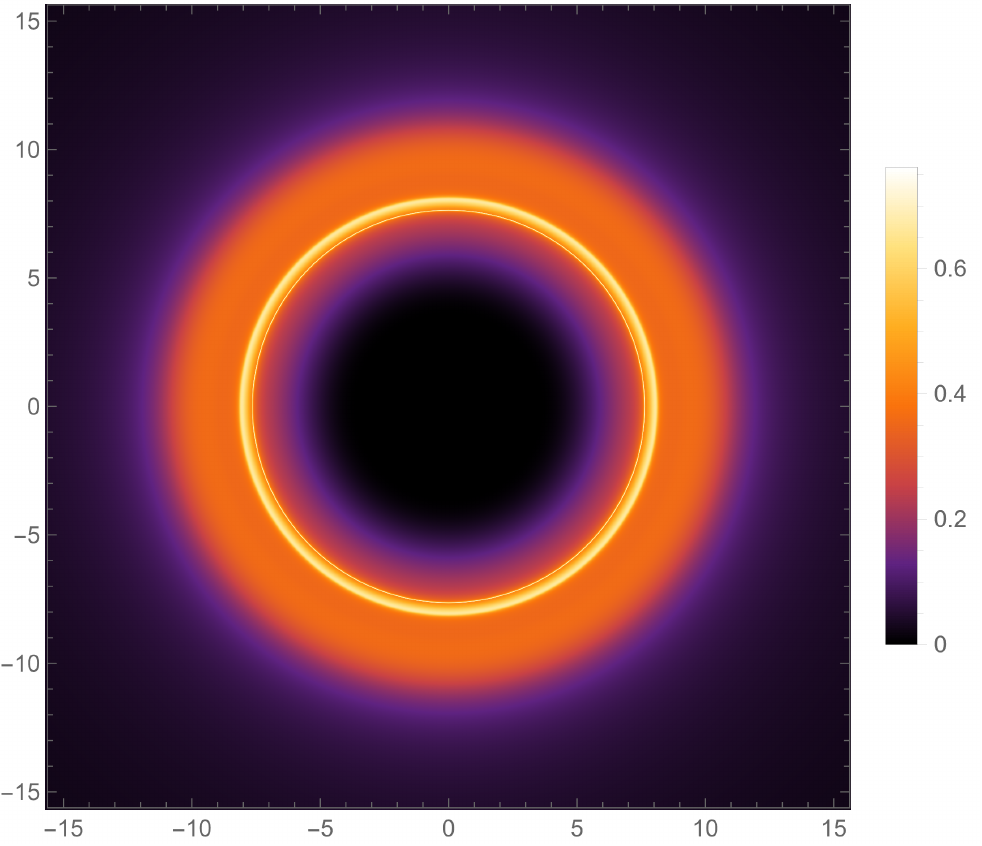}
 		\includegraphics[width=3.5cm]{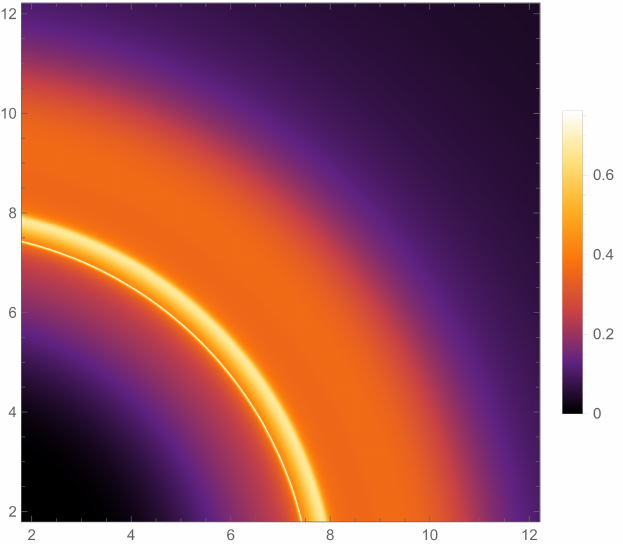}}\\
 	\subfigure[\, $b_2=8$]
 	{\includegraphics[width=4.2cm]{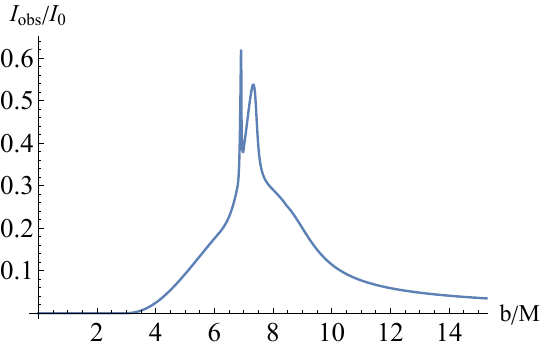} \label{}\hspace{2mm} \includegraphics[width=4.2cm]{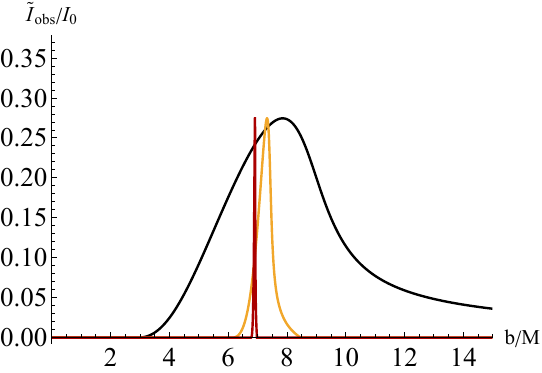}\hspace{2mm} \includegraphics[width=3.6cm]{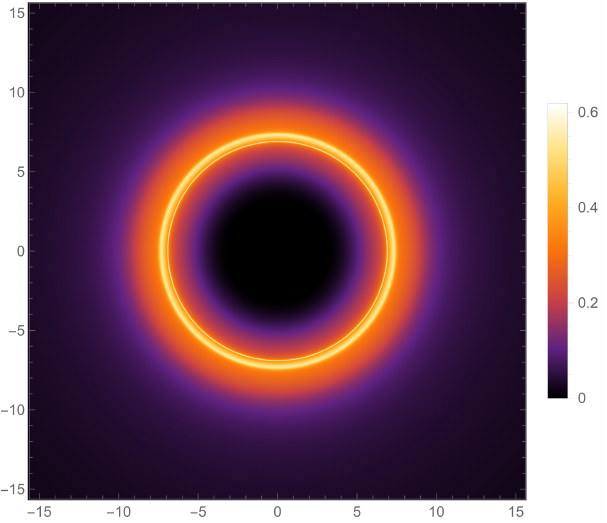}
 		\includegraphics[width=3.5cm]{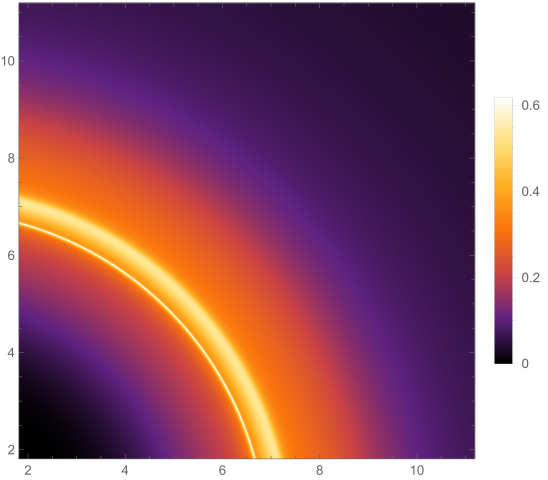}}\\
 	\subfigure[\, $b_2=9.5$]
 	{\includegraphics[width=4.2cm]{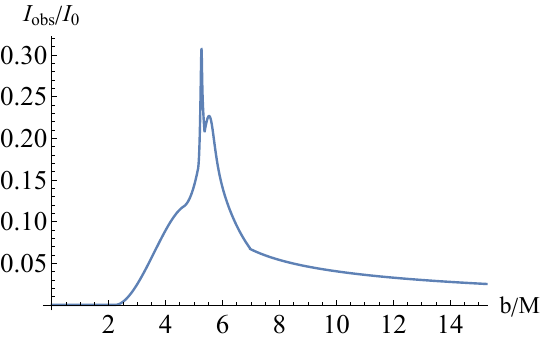} \label{}\hspace{2mm} \includegraphics[width=4.2cm]{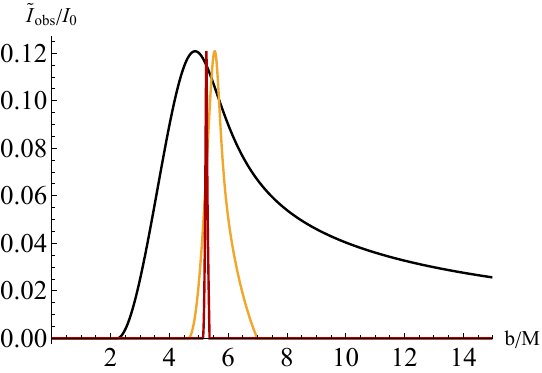}\hspace{2mm} \includegraphics[width=3.6cm]{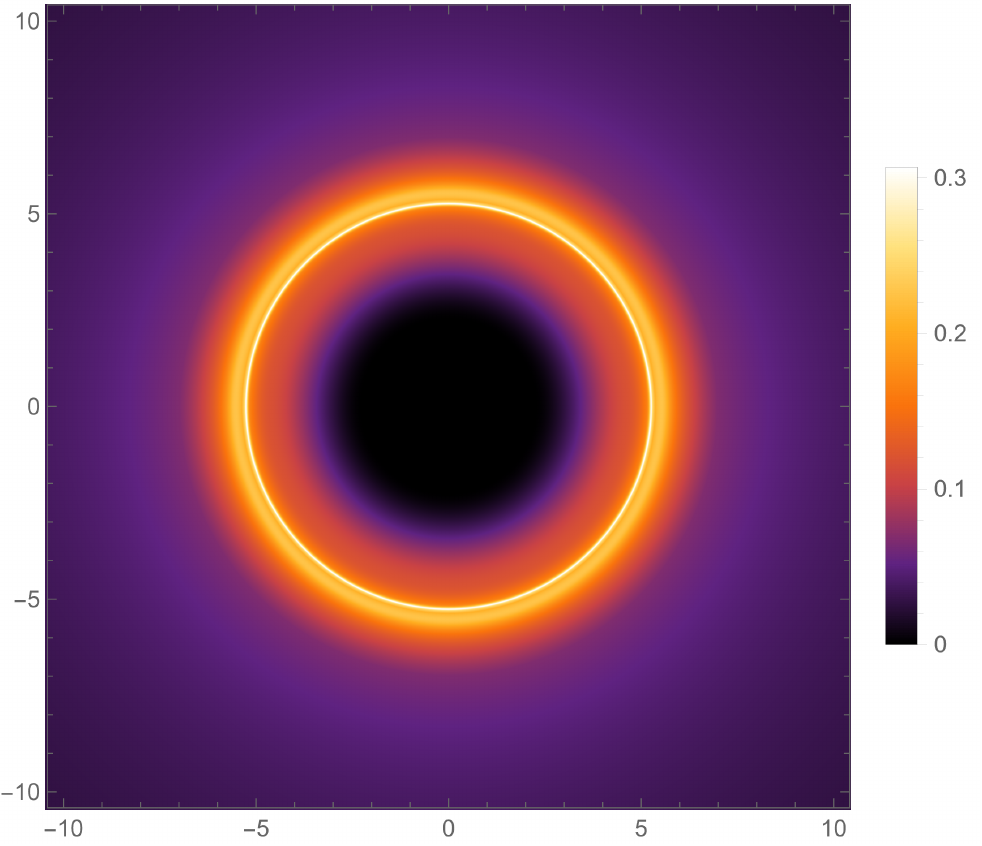}
 		\includegraphics[width=3.5cm]{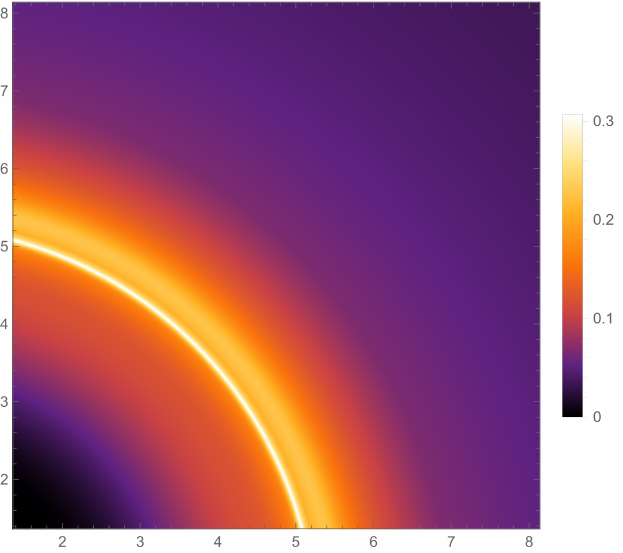}}\\
 	\caption{Observational appearances of the thin accretion disk with intensity profile II in \eqref{Profile-V3}, for various values of $b_2$ and $a_2=8$. The first column shows the total observed intensities $I_{obs}/I_0$ as a function of the impact parameter $b$. The second column displays the individual observed intensities corresponding to the first (black), second (yellow), and third (red) transfer functions in Eq.~\eqref{IOBS-V2}. The third column depicts the optical appearances, illustrating the distribution of observed intensities over two-dimensional plane. The fourth column provides zoomed-in views of these optical appearance.}
 	\label{figprofile3fulu}
 \end{figure}

Fig.\ref{figprofile3fulu} illustrates the observational appearances illuminated by the thin accretion disk II, as characterized by the emitted intensity profile II given in Eq.\eqref{Profile-V3}. The second column shows that the direct intensity has a relatively broad distribution over the impact parameter $b$, differing significantly from that illuminated by the thin accretion disk I. As the quantum parameter $b_2$ increases, the direct, lensed and photon ring intensity distributions shift to the left.
However, the direct intensity distribution moves more slowly relative to the lensed and photon ring distributions. Consequently, as $b_2$ increases, the peaks of the lensed and photon ring distributions, initially to the left of the direct intensity peak, eventually shift to the right of it. Due to these distinct differences in the intensity distributions compared to those illuminated by the thin accretion disk I, the total observed intensities and two-dimensional optical images, shown in the first, third, and fourth columns of Fig.\ref{figprofile3fulu}, exhibit unique features that differ from those predicted for a Schwarzschild BH.

\subsection{Thin disk accretion III}\label{appendix-C}

Furthermore, we investigate the characteristics of the observational appearances illuminated by the thin accretion disk, assuming that the radiation originates from the photon sphere. The intensity of this radiation, denoted as profile III, is given by
\begin{eqnarray}
	\text{profile III}: I_{em}(r)=\begin{cases}
		I_0[\frac{1}{r-(r_{ph}-1)}]^3	& \text{ if } r>r_{ph} \\
		0	& \text{ if } r\leq r_{ph} 
	\end{cases}\,,
	\label{Profile-V2fulu}
\end{eqnarray}
where $r=r_{ph}$ is the radius of photon sphere.
\begin{figure}[htbp]
	\centering
	\subfigure[\, $b_2=5$]
	{\includegraphics[width=4.2cm]{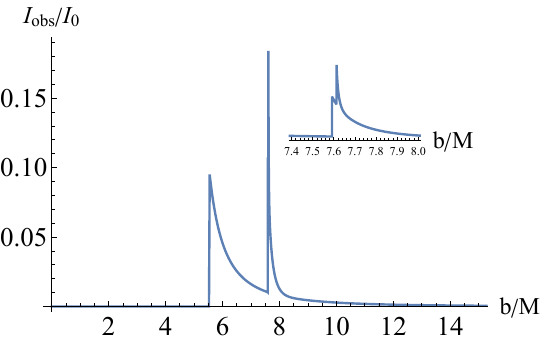} \label{}\hspace{2mm} \includegraphics[width=4.2cm]{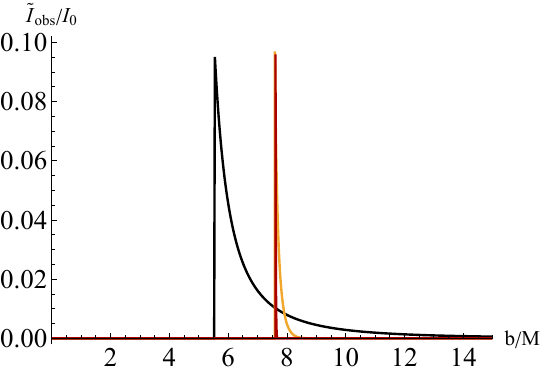}\hspace{2mm} \includegraphics[width=3.6cm]{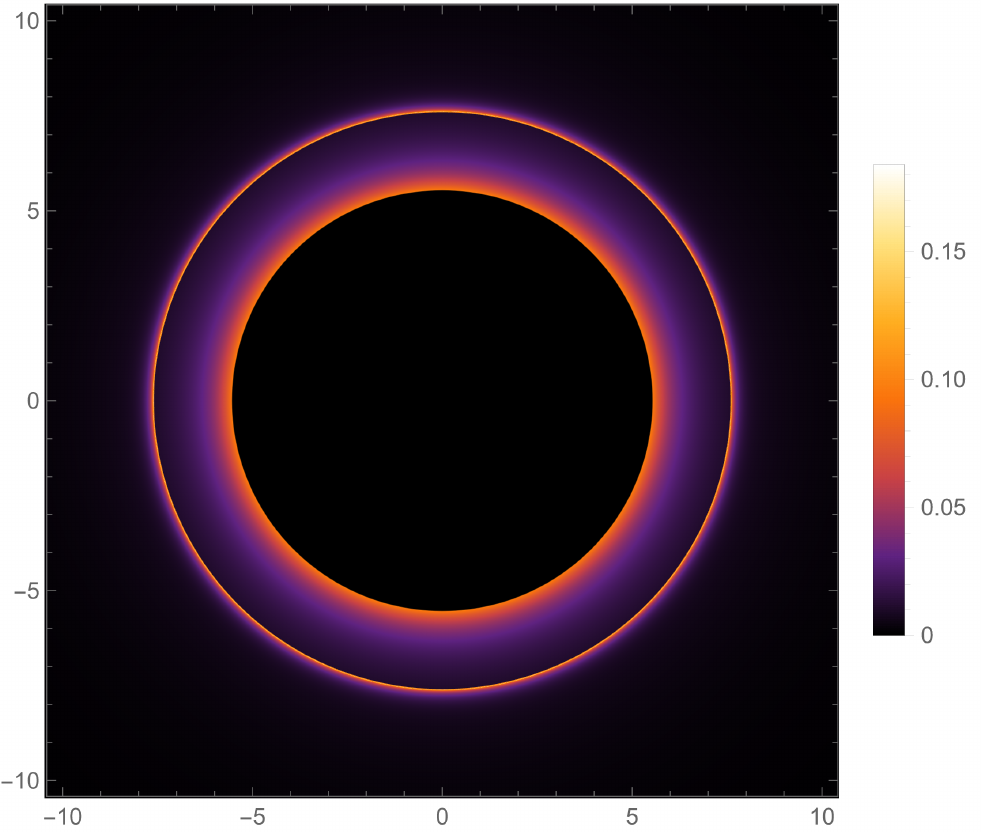}
		\includegraphics[width=3.5cm]{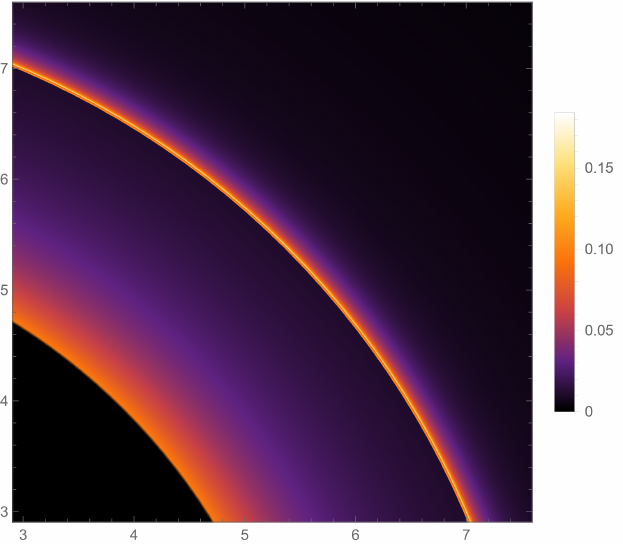}}\\
	\subfigure[\, $b_2=8$]
	{\includegraphics[width=4.2cm]{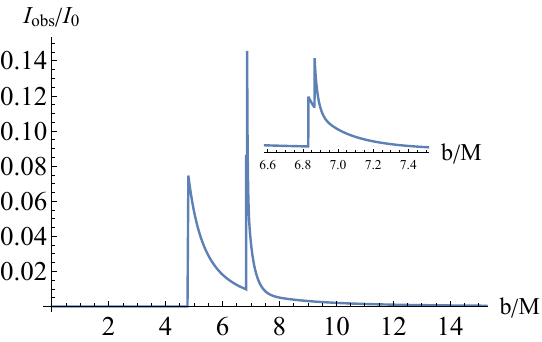} \label{}\hspace{2mm} \includegraphics[width=4.2cm]{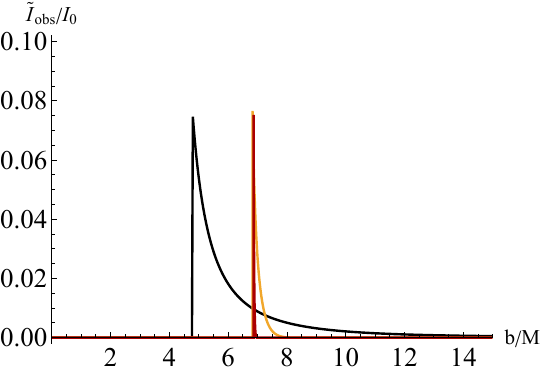}\hspace{2mm} \includegraphics[width=3.6cm]{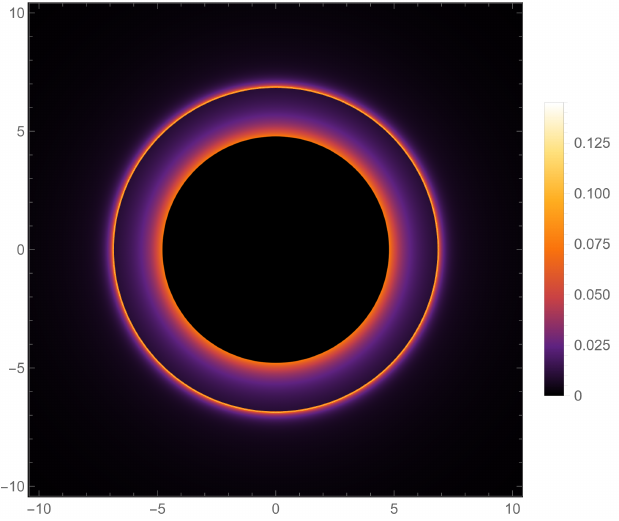}
		\includegraphics[width=3.5cm]{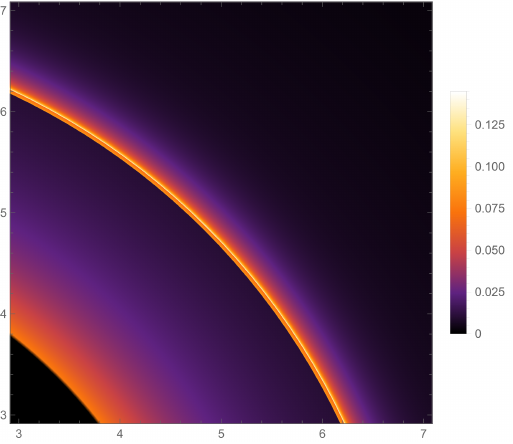}}\\
	\subfigure[\, $b_2=9.5$]
	{\includegraphics[width=4.2cm]{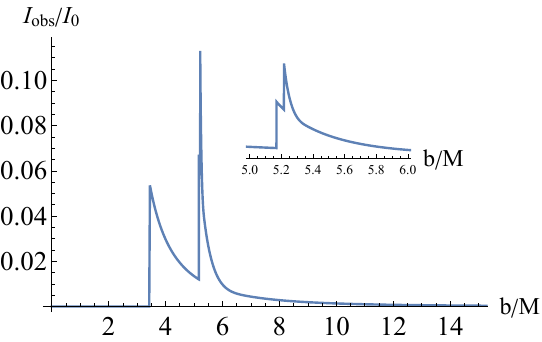} \label{}\hspace{2mm} \includegraphics[width=4.2cm]{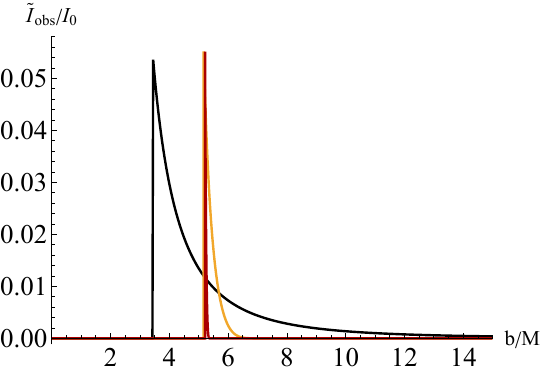}\hspace{2mm} \includegraphics[width=3.6cm]{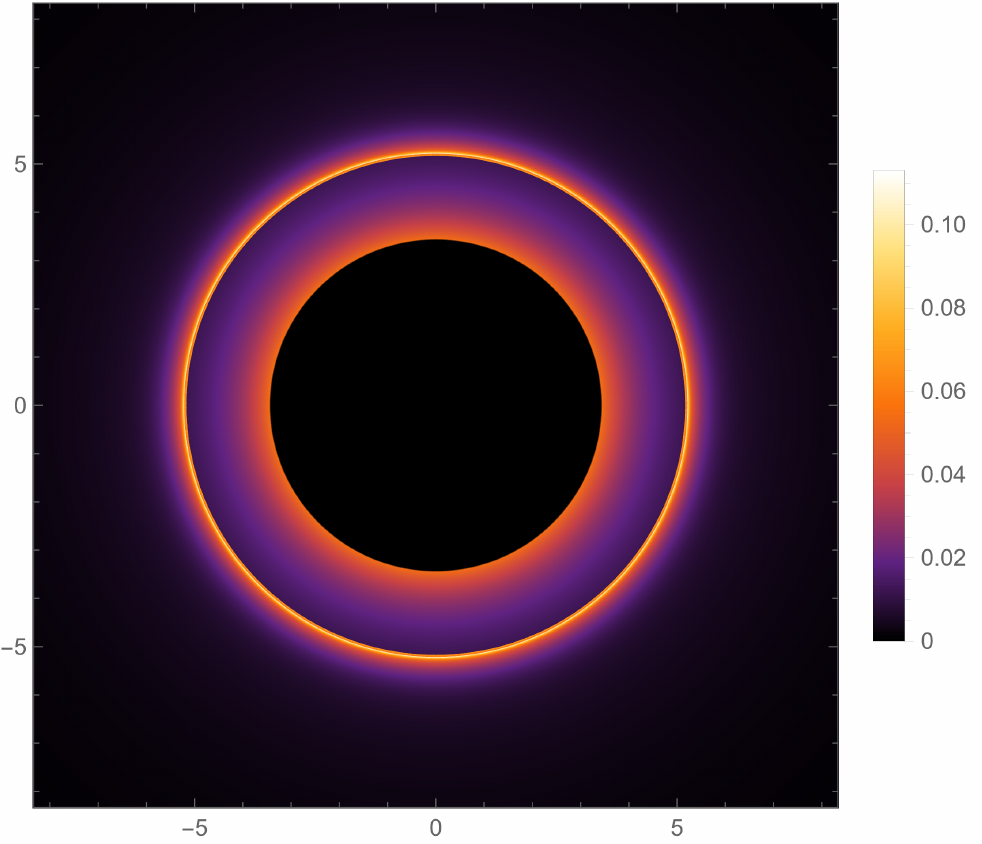}
		\includegraphics[width=3.5cm]{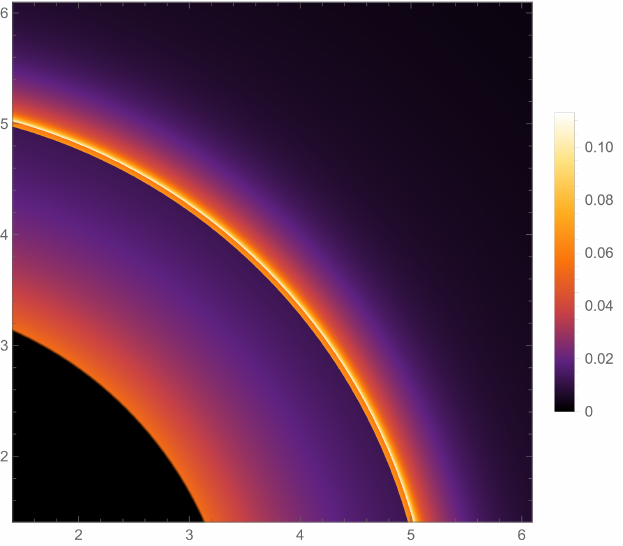}}\\
	\caption{Observational appearances of the thin accretion disk with intensity profile III in \eqref{Profile-V2fulu}, for various values of $b_2$ and $a_2=8$. The first column shows the total observed intensities $I_{obs}/I_0$ as a function of the impact parameter $b$. The second column displays the individual observed intensities corresponding to the first (black), second (yellow), and third (red) transfer functions in Eq.~\eqref{IOBS-V2}. The third column depicts the optical appearances, illustrating the distribution of observed intensities over two-dimensional plane. The fourth column provides zoomed-in views of these optical appearance.
    }
	\label{figprofile2fulu}
\end{figure}

The optical features of this regular BH illuminated by this accretion disk are illustrated in Fig.\ref{figprofile2fulu}. The main characteristics can be summarized as follows:
\begin{itemize}
    \item As the quantum parameter $b_2$ increases, the observed intensity shifts to the left, a behavior similar to that observed with thin accretion disks I and II.
    \item The second and third transfer functions broaden with increasing $b_2$, suggesting a greater contribution to the total observed intensity.
    \item The first three transfer functions are superimposed, with the direct intensity always located inside the lensing intensity and the photon ring intensity. This configuration results in a total observed intensity that differs from that of observations illuminated by thin accretion disks I and II. 
\end{itemize}

However, it is important to note that, unlike the results from the previous two accretion disks, the total observed intensity in this model shows minimal differences compared to the Schwarzschild BH case \cite{Gralla:2019xty}. This suggests that the quantum correction parameter $b_2$ does not significantly affect the observed intensity when the BH is illuminated by accretion disk III.

\bibliographystyle{style1}
\bibliography{Ref}
\end{document}